\documentclass[12pt]{article}
\usepackage{a4wide,latexsym,graphicx,epsfig,psfrag,float}
\usepackage{amsmath}
\usepackage{longtable}
\usepackage{amssymb}
\usepackage{bbm}
\usepackage{cite}
\allowdisplaybreaks

\newcommand{\lsim}{\stackrel{<}{_\sim}}

\def\mapright#1#2{\smash{
     \mathop{-\!\!\!-\!\!\!\rightarrow}\limits^{#1}_{#2}}}

\providecommand{\openone}{\leavevmode\hbox{\small1\kern-3.8pt\normalsize1}}

\def\mapright#1#2{\smash{
     \mathop{-\!\!\!-\!\!\!-\!\!\!\longrightarrow}\limits^{#1}_{#2}}}

\begin{document}
\parskip=3pt plus 1pt

\begin{titlepage}
\begin{flushright}
{
FTUV/09-1111 \\
IFIC/09-53 \\
LPT-Orsay-09-90
}
\end{flushright}
\vskip 3.5cm

\setcounter{footnote}{0}
\renewcommand{\thefootnote}{\fnsymbol{footnote}}

\begin{center}
{\LARGE \bf Hadron structure in  $\tau \rightarrow KK\pi \nu_\tau$ decays}
\\[35pt]

{ \sc  D.~G\'omez Dumm$^{1}$,
P.~Roig$^{2}$,
A.~Pich$^{3}$,
J. Portol\'es$^{3}$}

\vspace{1.5cm}
${}^{1)}$ IFLP, CONICET $-$ Dpto. de F\'{\i}sica,
Universidad Nacional de La Plata,  \\ C.C.\ 67, 1900 La Plata, Argentina\\[15pt] 
${}^{2)}$ Laboratoire de Physique Th\'eorique (UMR 8627), Universit\'e de Paris-Sud XI, \\ B\^{a}timent 210,
91405 Orsay cedex, France \\[15pt]
${}^{3)}$ Departament de F\'{\i}sica Te\`orica, IFIC, CSIC ---
Universitat de Val\`encia, \\
Edifici d'Instituts de Paterna, Apt. Correus 22085, E-46071
Val\`encia, Spain \\[10pt]
\end{center}

\setcounter{footnote}{0}
\renewcommand{\thefootnote}{\arabic{footnote}}

\vfill

\begin{abstract}
\noindent We analyse the hadronization structure of both vector and
axial-vector currents leading to $\tau\to KK\pi \,\nu_\tau$ decays.  At
leading order in the $1/N_C$ expansion, and considering only the
contribution of the lightest resonances, we work out, within the framework
of the resonance chiral Lagrangian, the structure of the local vertices
involved in those processes. The couplings in the resonance theory are
constrained by imposing the asymptotic behaviour of vector and axial-vector
spectral functions ruled by QCD. In this way we predict the hadron spectra
and conclude that, contrarily to previous assertions, the vector
contribution dominates by far over the axial-vector one in all $K K \pi$
charge channels.
\end{abstract}

PACS~: 11.15.Pg, 12.38.-t, 12.39.Fe \\ \hspace*{0.5cm}
Keywords~: Hadron tau decays, chiral Lagrangians, QCD, $1/N$ expansion.
\vfill

\end{titlepage}

\section{Introduction} \label{sect:1}

\hspace*{0.5cm} Hadron decays of the tau lepton provide a prime scenario
to study the hadronization of QCD currents in an energy region settled by
many resonances. This task has a twofold significance. First, the study of
branching fractions and spectra of those decays is a major goal of the
asymmetric B factories (BABAR, BELLE). These are supplying an
enormous amount of quality data owing to their large statistics, and the
same is planned for the near future at tau-charm factories such as BES-III.
Second, the required hadronization procedures involve QCD in a
non-perturbative energy region ($E \lsim M_{\tau} \sim 1.8 \, \mbox{GeV}$)
and, consequently, these processes are a clean benchmark, not spoiled by an
initial hadron state, where we can learn about the treatment of strong
interactions when driven by resonances.
\par
Analyses of tau decay data involve matrix elements that convey the
hadronization of the vector and axial-vector currents. At present there is
no determination from first principles of those matrix elements as they
involve strong interaction effects in its non-perturbative regime. Therefore we
have to rely in models that parameterize the form factors that arise from
the hadronization. A relevant one is the so-called K\"uhn-Santamar\'{\i}a model
(KS) \cite{Pich:1987qq} that, essentially, relies on the construction of
form factors in terms of Breit-Wigner functions weighted by unknown
parameters that are extracted from phenomenological analyses of data.
This procedure, that has proven to be successful in the description of the $\pi
\pi \pi$ final state, has been employed in the study of many two- and
three-hadron tau decays \cite{KS3,KS5,KS6,Gomez-Cadenas:1990uj}. The ambiguity related with the
choice of Breit-Wigner functions \cite{Pich:1987qq,GS} is currently being
exploited to estimate the errors in the determination of the free
parameters. The measurement of the $K K \pi$ spectrum by the
CLEO Collaboration \cite{Liu:2002mn} has shown that the parameterization
described by the KS model does not recall appropriately the experimental features
keeping, at the same time, a consistency with the underlying strong
interaction theory \cite{Portoles:2004vr}. The solution provided by CLEO
based in the introduction of new parameters spoils the normalization
of the Wess-Zumino anomaly, i.e. a specific prediction of QCD.
Indeed, arbitrary
parameterizations are of little help in the procedure of obtaining
information about non-perturbative QCD. They may fit the data but do not
provide us hints on the hadronization procedures. The key point in order to
uncover the inner structure of hadronization is to guide the construction of
the relevant form factors with the use of known properties of QCD.
\par
The TAUOLA library \cite{tauola} is, at present, a key tool that handles
analyses of tau decay data. Though originally it comprehended assorted versions
of the KS model only, it has been opened to the introduction of matrix
elements obtained with other models. Hence it has become an excellent tool
where theoretical models confront experimental data. This or analogous
libraries are appropriate benchmarks where to apply the results of our research.
\par
 At very low energies ($E \ll M_{\rho}$, being $M_{\rho}$ the mass of the
$\rho(770)$ resonance) the chiral symmetry of massless QCD rules the
construction of an effective field theory that allows a perturbative
expansion in momenta ($p$) and light quark masses ($m$), 
as $(p^2, M_{\pi}^2) / \Lambda_{\chi}^2$, being
$\Lambda_{\chi} \sim 4 \pi F \sim M_{\rho}$ the scale that breaks the chiral
symmetry; here $M_{\pi}$ is the pion mass and $F$ is the decay
constant of the pion. Indeed Chiral Perturbation Theory ($\chi PT$)
\cite{Weinberg:1978kz} drives the hadronization of QCD currents into the
lightest multiplet of pseudoscalar mesons, $\pi$, $K$ and $\eta$. The
application of this framework to the study of pion decays of the tau lepton
was carried out in Ref.~\cite{Colangelo:1996hs} though, obviously, it can
only describe a tiny region of the available phase space. It is clear that,
whatever the structure given to the form factors in the region of
resonances, it should match the chiral constraint in its energy domain. In
fact the parameterizations with Breit-Wigner functions that are at the
center of the KS model fail to fulfill that condition already at ${\cal
O}(p^4)$ in the chiral expansion \cite{Portoles:2000sr,GomezDumm:2003ku}.
\par
Our knowledge of QCD in the energy region of the resonances is pretty poor.
Contrarily to the very low energy domain, we do not know how to construct a
dual effective field theory of strong interactions for $E \sim M_{\rho}$.
There is a tool, though, that could shed light on the appropriate structure
of a Lagrangian theory that we could use. This is yielded by the large-$N_C$
limit of $SU(N_C)$ QCD \cite{'tHooft:1973jz}, which introduces an expansion
in inverse powers of the number of colours $N_C$. The essential idea
relevant for our goal that comes out from that setting is that at leading
order in the expansion, i.e. $N_C \rightarrow \infty$, any amplitude is
given by the tree level diagrams generated by a local Lagrangian with an
spectrum of infinite zero-width states. This frame, as we will see, can be
used to establish a starting point in the study of the hadron resonance
region and, consequently, in the hadron decays of the tau lepton. The
setting recalls the role of the resonance chiral theory
\cite{Ecker:1988te,Donoghue:1988ed} that can be better understood in the
light of the large-$N_C$ limit \cite{Peris:1998nj,Pich:2002xy,Cirigliano:2006hb}.
\par
At high energies ($E \gg M_{\rho}$), where the light-flavoured continuum is
reached, perturbative QCD is the appropriate framework to  deal with the
description of strong interaction of partons. In particular, a well known
feature of form factors of QCD currents is their smooth behaviour at high
transfer of momenta \cite{Brodsky}, thus it is reasonable to expect that the
form factors match this behaviour above the energy region of the resonances.
Another related tool is the study of the Operator Product Expansion (OPE) of
Green functions of QCD currents that are order parameters of the chiral
symmetry breaking. It is possible to evaluate these Green functions within a
resonance theory, and then perform a matching with their leading term of the
OPE expansion at high transfers of
momenta~\cite{Moussallam:1997xx,Knecht:2001xc,Cirigliano:2004ue,RuizFemenia:2003hm,Cirigliano:2005xn,Mateu:2007tr,Ecker:1989yg}.
In general, the information coming from high energies is important to settle
a resonance Lagrangian. It is reasonable to assume that the effective
couplings collect information coming from the energy region above the
resonances, hence the described procedure should help to determine the
corresponding coupling constants. Indeed, this approach has proven to be
capable of that task \cite{Ecker:1989yg}.
\par
In Ref.~\cite{GomezDumm:2003ku} we considered all mentioned steps in order
to analyse the $\pi \pi \pi$ hadron final state in the decay of the tau
lepton. Here we continue that undertaking by considering the $K K \pi$
channels that, as mentioned above, do not fit well within the KS model and
the present TAUOLA setup. Contrarily to the $\pi \pi \pi$ final state,
which is dominated by the hadronization of the axial-vector current,
$\tau\to K K \pi\nu_\tau$ decays receive contributions from both vector and
axial-vector currents. Indeed, one of the goals of our work is to find out
the relative weight of those contributions. Fortunately we will be assisted
in this task by the recent analysis of $e^+ e^- \rightarrow K K \pi$
cross-section by BABAR \cite{Aubert:2007ym} where a separation between
isoscalar and isovector channels has been performed. Hence we will be able
to connect both processes through CVC.
\par
In Section~2 we introduce the observables to be considered and the framework
settled by the procedure sketched above. Then the amplitudes for $K K \pi$
decay channels are evaluated in Section~3. An analysis of how we can get
information on the resonance couplings appearing in the hadronization of the
currents is performed in Section~4. Finally we explain our results in
Section~5, and our conclusions are pointed out in Section~6. Four technical
appendices complete our exposition.

\section{Theoretical framework} \label{sect:2}

\hspace*{0.5cm}The hadronization of the currents that rule semileptonic tau
decays is driven by non-perturbative QCD. As mentioned in the Introduction,
our methodology stands on the construction of an action, with the relevant
degrees of freedom, led by the chiral symmetry and the known asymptotic
behaviour of form factors and Green functions driven by large $N_C$ QCD. We
will limit ourselves to those pieces of the action that are relevant for the
study of decays of the tau lepton into three pseudoscalar mesons. Hence we
will need to include both even- and odd-intrinsic parity sectors.
\par
The large $N_C$ expansion of $SU(N_C)$ QCD implies that, in the $N_C
\rightarrow \infty$ limit, the study of Green functions of QCD currents can
be carried out through the tree level diagrams of a Lagrangian theory that
includes an infinite spectrum of non-decaying states \cite{'tHooft:1973jz}.
Hence the study of the resonance energy region can be performed by
constructing such a Lagrangian theory. The problem is that we do not
know how to implement an infinite spectrum in a model-independent way.
However, it is well known from the phenomenology that the main role is
always played by the lightest resonances. Accordingly it was suggested in
Refs.~\cite{Ecker:1988te,Donoghue:1988ed} that one can construct  a
suitable effective Lagrangian involving the lightest nonets of resonances
and the octet of Goldstone bosons states ($\pi$, $K$ and $\eta$). This is
indeed an appropriate tool to handle the hadron decays of the tau lepton.
The guiding principle in the construction of such a Lagrangian is  
chiral symmetry. When
resonances are integrated out from the theory, i.e. one tries to describe
the energy region below such states ($E \ll M_{\rho}$), the remaining
setting is that of $\chi$PT, to which now we turn.
\par
The very low-energy strong interaction in the light quark sector is known to
be ruled by the $SU(3)_L\otimes SU(3)_R$ chiral symmetry of massless QCD
implemented in $\chi$PT. The leading even-intrinsic-parity ${\cal O}(p^2)$
Lagrangian, which carries the information of the spontaneous symmetry
breaking of the theory, is~:
\begin{equation} \label{eq:op2}
{\cal L}_{\chi {\rm PT}}^{(2)}=\frac{F^2}{4}\langle u_{\mu}
u^{\mu} + \chi _+ \rangle \ ,
\end{equation}
where
\begin{eqnarray}
u_{\mu} & = & i [ u^{\dagger}(\partial_{\mu}-i r_{\mu})u-
u(\partial_{\mu}-i \ell_{\mu})u^{\dagger} ] \ , \nonumber \\
\chi_{\pm} & = & u^{\dagger}\chi u^{\dagger}\pm u\chi^{\dagger} u\ \
\ \ , \ \ \ \
\chi=2B_0(s+ip) \; \; ,
\end{eqnarray}
and $\langle \ldots \rangle$ is short for a trace in the flavour space.
The Goldstone octet of pseudoscalar fields
\begin{equation}  \label{eq:phi_matrix}
\Phi(x) =
\left(
\begin{array}{ccc}
 \displaystyle\frac{1}{\sqrt 2}\,\pi^0 + \displaystyle\frac{1}{\sqrt
 6}\,\eta_8
& \pi^+ & K^+ \\
\pi^- & - \displaystyle\frac{1}{\sqrt 2}\,\pi^0 +
\displaystyle\frac{1}{\sqrt 6}\,\eta_8
& K^0 \\
 K^- & \bar{K}^0 & - \displaystyle\frac{2}{\sqrt 6}\,\eta_8
\end{array}
\right)
\ ,
\end{equation}
is realized non--linearly into the unitary matrix in the
flavour space
\begin{equation}
u(\varphi)=\exp \left\{ \frac{i}{\sqrt{2}\,F} \Phi(x) \right\} \; \; \; ,
\end{equation}
 which under chiral rotations transforms as
\begin{equation}
u(\varphi)  \to  g_R\, u(\varphi)\, h(g,\varphi)^\dagger
                 = h(g,\varphi)\, u(\varphi)\, g_L^\dagger \; \; ,
\end{equation}
with $g \equiv (g_L,g_R) \, \in \, SU(3)_{\mathrm{L}} \otimes
SU(3)_{\mathrm{R}}$ and $h(g,\varphi)\,\in \, SU(3)_V$. External hermitian
matrix fields $r_{\mu}$, $\ell_{\mu}$, $s$ and $p$ promote the global
$SU(3)_{\mathrm{L}} \otimes SU(3)_{\mathrm{R}}$ symmetry to a local one.
Thus, interactions with electroweak bosons can be accommodated through the
vector $v_{\mu} = (r_{\mu} + \ell_{\mu}) / 2$ and axial--vector $a_{\mu} =
(r_{\mu} - \ell_{\mu}) / 2$ fields. The scalar field $s$ incorporates
explicit chiral symmetry breaking through the quark masses taking $s = {\cal
M} \, + \ldots$, with ${\cal M} =  \mathrm{diag}(m_u,m_d,m_s)$ and, finally,
at lowest order in the chiral expansion $F = F_{\pi} = 92.4$~MeV is
the pion decay constant and $B_0 F^2 = - \langle 0 | \bar{\psi}\psi | 0
\rangle_0$.
\par
The leading action in the odd-intrinsic-parity sector arises at ${\cal O}(p^4)$.
This is given by the chiral anomaly \cite{Wess:1971yu}
and explicitly stated by the Wess-Zumino-Witten ${\cal Z}_{WZ}[v,a]$ functional that can be
read in Ref.~\cite{Ecker:1994gg}. This contains all anomalous contributions to electromagnetic
and semileptonic meson decays.
\par
It is well known \cite{Ecker:1988te,Cirigliano:2006hb} that higher orders in
the chiral expansion, i.e.\ even-intrinsic-parity ${\cal L}_{\chi PT}^{(n)}$
with $n>2$, bring in the information of heavier degrees of freedom that have
been integrated out, for instance resonance states. As our next step intends
to include the latter explicitly, to avoid double counting issues
we will not consider higher orders in $\chi$PT. As we comment below, in
order to fulfill this procedure ---at least, up to ${\cal O}(p^4)$---
it is convenient to use the antisymmetric tensor representation for the
$J=1$ fields. Analogously, additional odd-intrinsic-parity amplitudes arise
at ${\cal O}(p^6)$ in $\chi$PT, either from one-loop diagrams using
one vertex from the Wess-Zumino-Witten action or from tree-level
operators \cite{Bijnens:2001bb}. However we will assume that the latter are
fully generated by resonance contributions \cite{RuizFemenia:2003hm} and,
therefore, will not be included in the following.
\par
The formulation of a Lagrangian theory that includes both the octet of
Goldstone mesons and $U(3)$ nonets of resonances is carried out through
the construction of a phenomenological Lagrangian \cite{Coleman:1969sm} where
chiral symmetry determines the 
structure of the operators. Given the vector character of the Standard Model
(SM) couplings of the hadron matrix elements in $\tau$ decays, form factors
for these processes are ruled by vector and
axial-vector resonances. Notwithstanding those form factors are given,
in the $\tau \rightarrow PPP \nu_{\tau}$ decays, by a
four-point Green function where other quantum numbers might play a role, 
namely scalar and pseudoscalar resonances \cite{Jamin:2000wn}. However their
contribution should be minor for $\tau \rightarrow K K \pi \nu_{\tau}$. 
Indeed the lightest scalar\footnote{As we assume the
$N_C \rightarrow \infty$ limit, the nonet of scalars corresponding to the
$f_0(600)$ is not considered. This multiplet is generated by 
rescattering of the ligthest pseudoscalars and then subleading in the $1/N_C$ expansion.},
namely $f_0(980)$, 
couples dominantly to two pions, and therefore its role in the $K \overline{K} \pi$ 
final state should be negligible.
Heavier flavoured or unflavoured scalars and pseudoscalars are at least suppressed
by their masses, being
heavier than the axial-vector meson $a_1(1260)$ 
(like $K_0^*(1430)$ that couples to $K \pi$). In addition the couplings of 
unflavoured states to
$K \overline{K}$ (scalars) and $K \overline{K} \pi$ (pseudoscalars) seem to be very
small \cite{PDG2008}.
Thus in our description we include $J=1$ resonances only \footnote{If
the study of these processes requires a more accurate description,
additional resonances could also be included in our scheme.}, and this is
done by considering a nonet of fields \cite{Ecker:1988te}~:
\begin{equation}
 R \,\equiv \, \frac{1}{\sqrt{2}} \, \sum_{i=1}^{9} \lambda_i \, \phi_{R,i}\; ,
\end{equation}
where $R = V, A$, stands for vector and axial-vector resonance states. Under
the $SU(3)_L \otimes SU(3)_R$ chiral group, $R$ transforms as~:
\begin{equation} \label{eq:rtrans}
 R \, \rightarrow \, h(g,\varphi) \, R \, h(g,\varphi)^{\dagger} \; .
\end{equation}
The flavour structure of the resonances is analogous to that of the
Goldstone bosons in Eq.~(\ref{eq:phi_matrix}). We also introduce the
covariant derivative
\begin{eqnarray}
\nabla_\mu X &  \equiv &     \partial_{\mu} X    + [\Gamma_{\mu}, X] \; \; , \\
\Gamma_\mu & = & \frac{1}{2} \, [ \,
u^\dagger (\partial_\mu - i r_{\mu}) u +
u (\partial_\mu - i \ell_{\mu}) u^\dagger \,] \; \; ,\nonumber
\end{eqnarray}
acting on any object $X$ that transforms as $R$ in Eq.~(\ref{eq:rtrans}),
like $u_{\mu}$ and $\chi_{\pm}$. The kinetic terms for the spin 1 resonances
in the Lagrangian read~:
\begin{equation} \label{eq:lag0}
{\cal L}_{ \rm kin}^R = -\frac{1}{2} \langle \,
\nabla^\lambda R_{\lambda\mu} \nabla_\nu R^{\nu\mu} \, \rangle + \frac{M_R^2}{4} \, \langle \,
R_{\mu\nu} R^{\mu\nu}\,  \rangle \; \; \; , \; \; \;  \; \; R \, = \, V,A \; ,
\end{equation}
$M_V$, $M_A$ being the masses of the nonets of vector and axial--vector
resonances in the chiral and large-$N_C$ limits, respectively.
Notice that we describe the resonance fields
through the antisymmetric tensor representation. With this description
one is able to collect, upon integration of resonances, the bulk of the
low-energy couplings at ${\cal O}(p^4)$ in $\chi$PT without the inclusion of
additional local terms \cite{Ecker:1989yg}. In fact it is necessary
to use this representation if one does not include the ${\cal L}_{\chi
PT}^{(4)}$ in the Lagrangian theory. Though analogous studies at higher
chiral orders have not been carried out, we will assume that no ${\cal
L}_{\chi PT}^{(n)}$ with $n=4,6,...$ in the even-intrinsic-parity and
$n=6,8,...$ in the odd-intrinsic-parity sectors need to be included in the
theory.
\par
The construction of the interaction terms involving resonance and Goldstone fields is driven by
chiral and discrete symmetries with a generic structure given by~:
\begin{equation}
 {\cal O}_i \, \sim \, \langle \, R_1 R_2 ... R_j \, \chi^{(n)}(\varphi) \, \rangle \, ,
\end{equation}
where $\chi^{(n)}(\varphi)$ is a chiral tensor that includes only
Goldstone and auxiliary fields. It transforms like $R$ in
Eq.~(\ref{eq:rtrans}) and has chiral counting $n$ in the frame of $\chi$PT.
This counting is relevant in the setting of the theory because, though the resonance
theory itself has no perturbative expansion, higher values of $n$ may originate
violations of the proper
asymptotic behaviour of form factors or Green functions. As a guide
we will include at least those operators that, contributing to our processes,
are leading when integrating out the resonances. In addition we do not include
operators with higher-order chiral tensors, $\chi^{(n)}(\varphi)$, that would
violate the QCD asymptotic behaviour
unless their couplings are severely fine tuned to ensure the needed 
cancellations of large momenta. In the odd-intrinsic-parity
sector, that gives the vector form factor, this amounts to include all
$\langle R \chi^{(4)} \rangle$ and $\langle RR \chi^{(2)} \rangle$ terms.  
In the even-intrinsic-parity couplings, giving the axial-vector form factors,
these are the terms $\langle R \chi^{(2)} \rangle$. However previous analyses
of the axial-vector contributions
\cite{GomezDumm:2003ku, Cirigliano:2004ue} show the relevant role of the
$\langle RR \chi^{(2)} \rangle$ terms that, accordingly, are also considered
here~\footnote{Operators $\langle R \chi^{(4)} \rangle$ that are non-leading
and have a worse high-energy behaviour, are not included in the even-intrinsic-parity
contributions as they have not played any role in previous related analyses.}.
\par
 We also assume exact $SU(3)$ symmetry in the construction of the interacting
terms, i.e. at level of couplings. Deviations from exact symmetry in hadronic
tau decays have been considered in Ref.~\cite{Moussallam:2007qc}. However we do not
include $SU(3)$ breaking couplings because we are neither considering 
next-to-leading corrections in the $1/N_C$ expansion.
\par
The lowest order interaction operators linear in the resonance fields have the structure
$\langle R \chi^{(2)}(\varphi)\rangle$. There are no odd-intrinsic-parity terms of this form.
The even-intrinsic-parity Lagrangian includes three coupling
constants \cite{Ecker:1988te}~:
\begin{eqnarray} \label{eq:lag1}
{\cal L}_2^{\mbox{\tiny V}} & = &  \frac{F_V}{2\sqrt{2}} \langle V_{\mu\nu}
f_+^{\mu\nu}\rangle + i\,\frac{G_V}{\sqrt{2}} \langle V_{\mu\nu} u^\mu
u^\nu\rangle  \; \, , \nonumber \\
{\cal L}_2^{\mbox{\tiny A}} & = &  \frac{F_A}{2\sqrt{2}} \langle A_{\mu\nu}
f_-^{\mu\nu}\rangle \;,
\end{eqnarray}
where
$f_\pm^{\mu\nu}  =  u F_L^{\mu\nu} u^\dagger \pm u^\dagger F_R^{\mu\nu}
u$ and $F_{R,L}^{\mu \nu}$ are the field strength tensors associated
with the external right- and left-handed auxiliary fields. All coupling parameters
$F_V$, $G_V$ and $F_A$ are real.
\par
The leading odd-intrinsic-parity operators, linear in the resonance fields, have
the form $\langle R \chi^{(4)}(\varphi)\rangle$. We will need those pieces that
generate~: i) the vertex with one vector resonance and three pseudoscalar fields;
ii) the vertex with one vector resonance, a vector current and one pseudoscalar. The
minimal Lagrangian with these features is~:
\begin{equation}
\label{eq:l4odd}
{\cal L}_4^{\mbox{\tiny V}} = \sum_{i=1}^5 \, \frac{g_i}{M_V} \, {\cal O}^i_{\mbox{\tiny VPPP}}
\; + \;
\sum_{i=1}^{7} \,\frac{c_i}{M_{V}}\,{\cal O}^i_{\mbox{\tiny{VJP}}} \, ,
\end{equation}
where $g_i$ and $c_i$ are real adimensional couplings, and the operators
read
\begin{itemize}
\item[1/] VPPP terms
\begin{eqnarray}
\label{eq:omegappp}
{\cal O}_{\mbox{\tiny VPPP}}^1 & = & i \, \varepsilon_{\mu\nu\alpha\beta} \, \left\langle
V^{\mu\nu} \, \left( \, h^{\alpha\gamma} u_{\gamma} u^{\beta} - u^{\beta} u_{\gamma}
h^{\alpha\gamma} \right) \right\rangle \, , \nonumber  \\ [2mm]
{\cal O}_{\mbox{\tiny VPPP}}^2 & = & i \, \varepsilon_{\mu\nu\alpha\beta} \, \left\langle
V^{\mu\nu} \, \left( \, h^{\alpha\gamma} u^{\beta} u_{\gamma} - u_{\gamma} u^{\beta}
h^{\alpha\gamma} \, \right) \right\rangle \, , \nonumber \\ [2mm]
{\cal O}_{\mbox{\tiny VPPP}}^3 & = & i \, \varepsilon_{\mu\nu\alpha\beta} \, \left\langle
V^{\mu\nu} \, \left( \, u_{\gamma} h^{\alpha\gamma} u^{\beta}  -  u^{\beta}
h^{\alpha\gamma} u_{\gamma} \, \right) \right\rangle \, , \nonumber \\ [2mm]
{\cal O}_{\mbox{\tiny VPPP}}^4 & = &  \varepsilon_{\mu\nu\alpha\beta} \, \left\langle
\left\lbrace  \,V^{\mu\nu} \,,\,  u^{\alpha}\, u^{\beta}\, \right\rbrace \,{\cal \chi}_{-} \right\rangle \, , \nonumber \\ [2mm]
{\cal O}_{\mbox{\tiny VPPP}}^5 & = &  \varepsilon_{\mu\nu\alpha\beta} \, \left\langle
 \,u^{\alpha}\,V^{\mu\nu} \, u^{\beta}\, {\cal \chi}_{-} \right\rangle \, ,
\label{eq:VPPP}
\end{eqnarray}
with $h_{\mu \nu} = \nabla_{\mu} u_{\nu} + \nabla_{\nu} u_{\mu}$, and
\item[2/] VJP terms \cite{RuizFemenia:2003hm}
\begin{eqnarray}
 {\cal O}_{\mbox{\tiny VJP}}^1 & = & \varepsilon_{\mu\nu\rho\sigma}\,
\langle \, \{V^{\mu\nu},f_{+}^{\rho\alpha}\} \nabla_{\alpha}u^{\sigma}\,\rangle
\; \; , \nonumber\\[2mm]
{\cal O}_{\mbox{\tiny VJP}}^2 & = & \varepsilon_{\mu\nu\rho\sigma}\,
\langle \, \{V^{\mu\alpha},f_{+}^{\rho\sigma}\} \nabla_{\alpha}u^{\nu}\,\rangle
\; \; , \nonumber\\[2mm]
{\cal O}_{\mbox{\tiny VJP}}^3 & = & i\,\varepsilon_{\mu\nu\rho\sigma}\,
\langle \, \{V^{\mu\nu},f_{+}^{\rho\sigma}\}\, \chi_{-}\,\rangle
\; \; , \nonumber\\[2mm]
{\cal O}_{\mbox{\tiny VJP}}^4 & = & i\,\varepsilon_{\mu\nu\rho\sigma}\,
\langle \, V^{\mu\nu}\,[\,f_{-}^{\rho\sigma}, \chi_{+}]\,\rangle
\; \; , \nonumber\\[2mm]
{\cal O}_{\mbox{\tiny VJP}}^5 & = & \varepsilon_{\mu\nu\rho\sigma}\,
\langle \, \{\nabla_{\alpha}V^{\mu\nu},f_{+}^{\rho\alpha}\} u^{\sigma}\,\rangle
\; \; ,\nonumber\\[2mm]
{\cal O}_{\mbox{\tiny VJP}}^6 & = & \varepsilon_{\mu\nu\rho\sigma}\,
\langle \, \{\nabla_{\alpha}V^{\mu\alpha},f_{+}^{\rho\sigma}\} u^{\nu}\,\rangle
\; \; , \nonumber\\[2mm]
{\cal O}_{\mbox{\tiny VJP}}^7 & = & \varepsilon_{\mu\nu\rho\sigma}\,
\langle \, \{\nabla^{\sigma}V^{\mu\nu},f_{+}^{\rho\alpha}\} u_{\alpha}\,\rangle
\;\; .
\label{eq:VJP}
\end{eqnarray}
\end{itemize}
Notice that we do not include analogous pieces with an axial-vector resonance, that
would contribute to the hadronization of the axial-vector current. This
has been thoroughly studied in Ref.~\cite{GomezDumm:2003ku} (see also Ref.~\cite{shortp})
in the description of the $\tau \rightarrow \pi \pi \pi \nu_{\tau}$ process and it is shown
that no $\langle A \chi^{(4)}(\varphi) \rangle$ operators are needed to describe its hadronization.
Therefore those operators are not included in our minimal description of
the relevant form factors.
\par
In order to study tau decay processes with three pseudoscalar mesons in the
final state one also has to consider non-linear terms in the resonance
fields. Indeed the hadron final state in $\tau \rightarrow PPP
\nu_{\tau}$ decays can be driven by vertices involving two resonances and a
pseudoscalar meson. The structure of
the operators that give those vertices is $\langle R_1 R_2
\chi^{(2)}(\varphi) \rangle$, and has been worked out before
\cite{GomezDumm:2003ku,RuizFemenia:2003hm}. They include both even- and
odd-intrinsic-parity terms~:
\begin{equation}
\label{eq:lag21}
{\cal L}_2^{\mbox{\tiny RR}} \, = \, \sum_{i=1}^{5} \, \lambda_i \,
{\cal O}^i_{\mbox{\tiny VAP}} \;
\; + \;
\sum_{i=1}^{4} \,d_i\,{\cal O}^i_{\mbox{\tiny{VVP}}}\; ,
\end{equation}
where $\lambda_i$, and $d_i$ are unknown real adimensional couplings.
The operators ${\cal O}^i_{\mbox{\tiny RRP}}$ are given by~:
\begin{itemize}
\item[1/] VAP terms
\begin{eqnarray}
\label{eq:VAP}
{\cal O}^1_{\mbox{\tiny VAP}} &  = & \langle \,  [ \, V^{\mu\nu} \, , \,
A_{\mu\nu} \, ] \,  \chi_- \, \rangle \; \; , \nonumber \\ [2mm]
{\cal O}^2_{\mbox{\tiny VAP}} & = & i\,\langle \, [ \, V^{\mu\nu} \, , \,
A_{\nu\alpha} \, ] \, h_\mu^{\;\alpha} \, \rangle \; \; , \\ [2mm]
{\cal O}^3_{\mbox{\tiny VAP}} & = &  i \,\langle \, [ \, \nabla^\mu V_{\mu\nu} \, , \,
A^{\nu\alpha}\, ] \, u_\alpha \, \rangle \; \; ,  \nonumber \\ [2mm]
{\cal O}^4_{\mbox{\tiny VAP}} & = & i\,\langle \, [ \, \nabla^\alpha V_{\mu\nu} \, , \,
A_\alpha^{\;\nu} \, ] \,  u^\mu \, \rangle \; \; , \nonumber \\ [2mm]
{\cal O}^5_{\mbox{\tiny VAP}} & =  & i \,\langle \, [ \, \nabla^\alpha V_{\mu\nu} \, , \,
A^{\mu\nu} \, ] \, u_\alpha \, \rangle \nonumber \; \; .
\end{eqnarray}
\item[2/] VVP terms
\begin{eqnarray}
{\cal O}_{\mbox{\tiny VVP}}^1 & = & \varepsilon_{\mu\nu\rho\sigma}\,
\langle \, \{V^{\mu\nu},V^{\rho\alpha}\} \nabla_{\alpha}u^{\sigma}\,\rangle
\; \; , \nonumber\\[2mm]
{\cal O}_{\mbox{\tiny VVP}}^2 & = & i\,\varepsilon_{\mu\nu\rho\sigma}\,
\langle \, \{V^{\mu\nu},V^{\rho\sigma}\}\, \chi_{-}\,\rangle
\; \; , \nonumber\\[2mm]
{\cal O}_{\mbox{\tiny VVP}}^3 & = & \varepsilon_{\mu\nu\rho\sigma}\,
\langle \, \{\nabla_{\alpha}V^{\mu\nu},V^{\rho\alpha}\} u^{\sigma}\,\rangle
\; \; , \nonumber\\[2mm]
{\cal O}_{\mbox{\tiny VVP}}^4 & = & \varepsilon_{\mu\nu\rho\sigma}\,
\langle \, \{\nabla^{\sigma}V^{\mu\nu},V^{\rho\alpha}\} u_{\alpha}\,\rangle
\; \; .
\label{eq:VVP}
\end{eqnarray}
\end{itemize}
We emphasize that ${\cal L}_2^{\mbox{\tiny RR}}$  is a complete basis for constructing vertices with only
one pseudoscalar meson; for a larger number of pseudoscalars additional operators
might be added. As we are only interested in tree-level diagrams, the equation of
motion arising from ${\cal L}_{\chi PT}^{(2)}$ in Eq.~(\ref{eq:op2}) has been used in
${\cal L}_4^{\mbox{\tiny V}}$ and ${\cal L}^{\mbox{\tiny RR}}_2$ to eliminate superfluous operators.
\par
Hence our theory is given by the Lagrangian~:
\begin{equation}
\label{eq:ourtheory}
 {\cal L}_{R \chi T} \, = \, {\cal L}_{\chi PT}^{(2)} \, + \, {\cal L }_{\mbox{\tiny kin}}^{\mbox{\tiny R}} \, + \,
{\cal L}_2^{\mbox{\tiny A}} \, + \,{\cal L}_2^{\mbox{\tiny V}} \, + \,
 {\cal L}_4^{\mbox{\tiny V}} \, + \, {\cal L}_2^{\mbox{\tiny RR}} \, .
\end{equation}
It is important to point out that the resonance theory constructed above is
not a theory of QCD for arbitrary values of the couplings in the interaction
terms. As we will see later on, these constants can be constrained by
imposing well accepted dynamical properties of the underlying theory.

\section{Vector and axial-vector currents in $\tau  \rightarrow  K  K \pi \, \nu_{\tau}$ decays}
\label{sect:3}

\hspace*{0.5cm} The decay amplitude for the $\tau \rightarrow KK
\pi\nu_{\tau}$ decays can be written in the Standard Model as
\begin{equation}
 {\cal M} \, = \, - \, \frac{G_F}{\sqrt{2}} \, V_{ud} \, \overline{u}_{\nu_{\tau}} \,
\gamma^{\mu}  \left( 1 \, - \, \gamma_5 \right) \, u_{\tau} \, T_{\mu} \; ,
\end{equation}
where $V_{ud}$ is an element of the Cabibbo-Kobayashi-Maskawa matrix and
$T_{\mu}$  is the hadron matrix element of the participating $V_{\mu} -
A_{\mu}$ QCD current~:
\begin{equation}
 T_{\mu} \, = \, \langle K(p_1) \,  K(p_2) \,  \pi(p_3) \, | \, \left( V_{\mu} - A_{\mu} \right)
e^{i {\cal L}_{QCD}} \, | \, 0 \rangle \, .
\end{equation}
The hadron tensor can be written in terms of four form factors $F_1$, $F_2$, $F_3$ and $F_4$ as
\cite{Kuhn:1992nz}~:
\begin{equation}
\label{eq:t3}
 T^{\mu} \, = \, V_1^{\mu} \, F_1 \, + \, V_2^{\mu} \, F_2 \, + \, V_3^{\mu} \, F_3 \,
+ \, Q^{\mu} \, F_4 \; ,
\end{equation}
where $Q^{\mu} = p_1^{\mu} + p_2^{\mu} + p_3^{\mu}$ and
\begin{eqnarray} \label{eq:v123}
 V_{1 \, \mu} &  = &  \left( \, g_{\mu \nu} - \frac{Q_{\mu} Q_{\nu}}{Q^2} \right)  \, \left( p_2 - p_1
\right)^{\nu} \;   , \; \; \;  \; \; \;  \; \; \;  \;
V_{2 \, \mu} \;  = \;  \left( \, g_{\mu \nu} - \frac{Q_{\mu} Q_{\nu}}{Q^2} \right)  \, \left( p_3 - p_1
\right)^{\nu} \, , \nonumber \\[5mm]
V_{3 \, \mu} &= & i \, \varepsilon_{\mu \nu \varrho \sigma} \, p_1^{\nu} \, p_2^{\varrho} \, p_3^{\sigma}
\;.
\end{eqnarray}
There are three different charge channels for the $KK\pi$ decays of the
$\tau^-$ lepton, namely $K^+ (p_+) \, K^-(p_-) \, \pi^-(p_{\pi})$, $K^0(p_0)
\, \overline{K}^0(\overline{p}_0) \, \pi^-(p_{\pi})$ and $K^-(p_-) \,
K^0(p_0) \, \pi^0(p_{\pi})$. The definitions of Eq.~(\ref{eq:v123})
correspond to the choice $p_3 = p_{\pi}$ in all cases, and~: $(p_1,p_2) =
(p_+,p_-)$ for the $K^+ \, K^-$ case, $(p_1,p_2) = (\overline{p}_0,p_0)$ for
$K^0 \, \overline{K}^0$  and $(p_1,p_2) = (p_-,p_0)$ for $K^- \, K^0$. In
general, form factors $F_i$ are functions of the kinematical invariants~:
$Q^2$, $s=(p_1+p_2)^2$ and $t=(p_1+p_3)^2$. $F_1$ and $F_2$ originate from
the axial-vector current, while $F_3$ follows from the vector current. All
of them correspond to spin-1 transitions. The $F_4$ pseudoscalar form factor
stems from the axial-vector current, and corresponds to a spin-0 transition.
It is seen that this form factor vanishes in the chiral limit, therefore its
contribution is expected to be heavily suppressed, and both the spectrum and
the branching ratio of tau decays into three pseudoscalar mesons is
dominated by $J=1$ transitions,  especially in the Cabibbo-allowed modes.
\par
The $Q^2$-spectrum is given by~:
\begin{equation}
 \frac{d \,\Gamma}{d \,Q^2} \, = \, \frac{G_F^2 \, |V_{ud}|^2}{128 \, (2 \pi)^5 \, M_{\tau}} \,
\left( \frac{M_{\tau}^2}{Q^2}-1 \right)^2 \, \int \, ds \, dt \,
\left[ W_{SA} + \frac{1}{3} \left(  1 + 2 \frac{Q^2}{M_{\tau}^2} \right) \left( W_A + W_B \right)
\right] \, ,
\end{equation}
where the hadron structure functions, introduced in Ref.~\cite{Kuhn:1992nz}, are~:
\begin{eqnarray}
W_A & = & - ( V_1^{\mu} F_1 + V_2^{\mu} F_2 ) ( V_{1\mu} F_1 + V_{2 \mu} F_2 )^* \, , \nonumber \\ [3mm]
W_B & = & \frac{1}{4} \left[ \,  s \, t \,  u \, + \,
(m_K^2 -m_{\pi}^2) \, ( Q^2-m_K^2) \, s \, + \, m_K^2 (2 m_{\pi}^2 - Q^2) \, Q^2
\ - \,  m_K^2 m_{\pi}^4 \, \right] \, | F_3 |^2 \, ,  \nonumber \\[3mm]
W_{SA} & = & (Q^{\mu} \, F_4) (Q_{\mu} F_4^*) \, = \, Q^2 \, |F_4|^2 \,,
\end{eqnarray}
where $u=Q^2 - s - t + 2 m_K^2 + m_{\pi}^2$. The phase-space integrals
extend over the region spanned by the hadron system with a center-of-mass
energy $\sqrt{Q^2}$~:
\begin{equation}
\label{eq:phaspace1}
 \int \, ds \, dt \, \equiv \, \int_{4 \, m_K^2}^{(\sqrt{Q^2}-m_{\pi})^2} ds \,
\int_{t_{-}(s)}^{t_+(s)} dt \,,
\end{equation}
with
\begin{equation}
\label{eq:phaspace2}
 t_{\pm}(s) \, = \, \frac{1}{4 \, s} \left\{ \left(Q^2-m_{\pi}^2 \right)^2 \, - \,
\left[ \lambda^{1/2}\left( Q^2, s, m_{\pi}^2 \right) \, \mp \, \lambda^{1/2} \left( m_K^2, m_K^2, s
\right) \right]^2 \right\} \, ,
\end{equation}
and $\lambda(a,b,c) = (a+b-c)^2-4ab$. We have neglected here the
$\nu_{\tau}$ mass, and exact isospin symmetry has been assumed.
\par
\begin{figure}
\begin{center}
\includegraphics[scale=0.9]{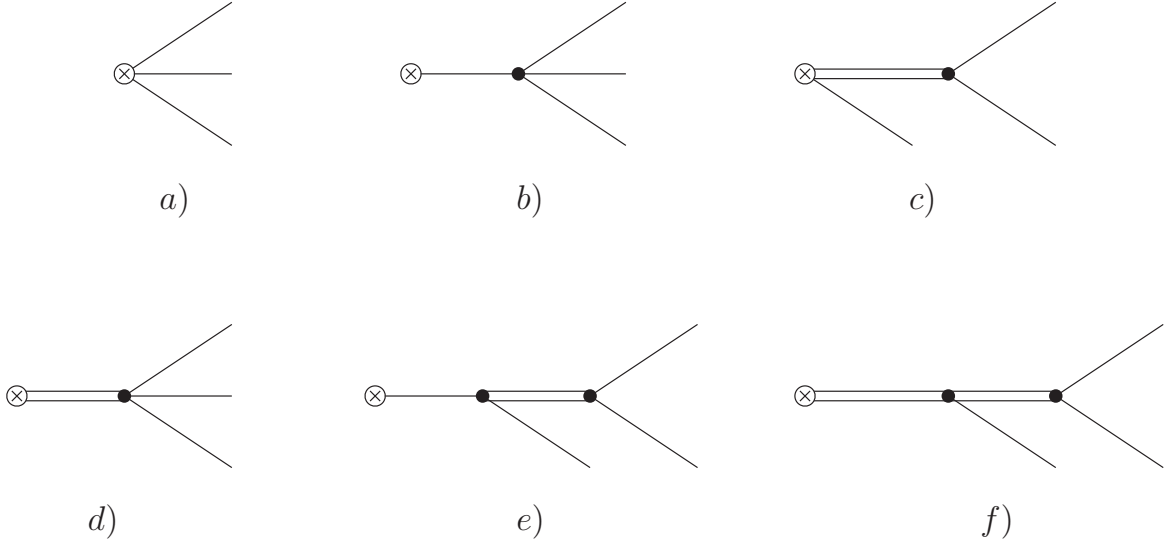}
\caption[]{\label{fig:feynman} Topologies contributing to the final hadron
state in $\tau \rightarrow K K \pi \, \nu_{\tau}$ decays in the $N_C
\rightarrow \infty$ limit. A crossed circle indicates the QCD vector or
axial-vector current insertion. A single line represents a pseudoscalar
meson ($K$, $\pi$)  while a double line stands for a resonance intermediate
state. Topologies $b)$ and $e) $ only contribute to the axial-vector driven
form factors, while diagram $d)$ arises only (as explained in the text)
from the vector current.}
\end{center}
\end{figure}

The general structure of the form factors, within our model, arises from the
diagrams displayed in Fig.~\ref{fig:feynman}.
This provides the following decomposition~:
\begin{equation}
 F_i \, = \, F_i^{\chi} \, + \, F_i^{\mbox{\tiny R}} \, + \, F_i^{\mbox{\tiny RR}} \;, \; \;  \, i=1,... \, ,
\end{equation}
where $F_i^{\chi}$ is given by the $\chi$PT Lagrangian [topologies $a)$ and
$b)$ in Fig.~\ref{fig:feynman}], and the rest are the contributions of one
[Fig.~1$c)$, $d)$ and $e)$] or two resonances [Fig.~1$f)$].

\subsection{Form factors in $\tau^-  \rightarrow  K^+ K^- \, \pi^- \,
\nu_{\tau}$ and $\tau^-  \rightarrow  K^0 \,\overline{K^0} \, \pi^- \,
\nu_{\tau}$}

\hspace*{0.5cm} In the isospin limit, form factors for the $\tau^-
\rightarrow K^+ K^- \pi^- \nu_{\tau}$ and $\tau^- \rightarrow K^0
\overline{K^0} \pi^- \nu_{\tau}$ decays are identical. The explicit
expressions for these are~:
\begin{eqnarray}
\label{eq:f11}
F_1^{\chi} &= & - \frac{\sqrt{2}}{3 \, F} \, , \nonumber \\
F_1^{\mbox{\tiny R}}(s,t) & = & - \, \frac{\sqrt{2}}{6} \, \frac{F_V\,G_V}{F^3}  \, \left[
\, \frac{A^{\mbox{\tiny R}}(Q^2,s,u,m_K^2,m_{\pi}^2,m_K^2)}{M_{\rho}^{2}-s}  \, +
\, \frac{B^{\mbox{\tiny R}}(s,u,m_K^2,m_{\pi}^2)}{M_{K^{*}}^{2}-t} \,\right] ,  \,  \\[4mm]
F_1^{\mbox{\tiny RR}}(s,t) & = & \frac{2}{3} \, \frac{F_A G_V}{F^3} \, \frac{Q^2}{M_{a_1}^2-Q^2} \,
\, \left[ \, \frac{A^{\mbox{\tiny RR}}(Q^2,s,u,m_K^2,m_{\pi}^2,m_K^2) }{M_{\rho}^2-s} \,
\right. \nonumber \\
& &  \qquad \qquad \qquad \qquad \qquad\left.
+ \, \frac{B^{\mbox{\tiny RR}}(Q^2,s,u,t,m_K^2,m_{\pi}^2,m_K^2)}{M_{K^*}^2-t} \, \right]
\, \, , \nonumber
\end{eqnarray}
where the functions $A^{\mbox{\tiny R}}$, $B^{\mbox{\tiny R}}$,
$A^{\mbox{\tiny RR}}$ and $B^{\mbox{\tiny RR}}$ are defined in
Appendix~\ref{ap:1}. The dependence of the form factors with $t$ follows
from the relation $u = Q^2 - s - t + 2 m_K^2 + m_\pi^2$. Moreover resonance
masses correspond to the lowest states, $M_{\rho} = M_{\rho (770)}$,
$M_{K^*} = M_{K^*(892)}$ and $M_{a_1} = M_{a_1(1260)}$\footnote{
Resonance masses and widths within our approach are discussed in
subsect.~3.3.}.
\par
Analogously the $F_2$ form factor is given by~:
\begin{eqnarray}
\label{eq:f21}
 F_2^{\chi} &= & F_1^{\chi} \, , \nonumber \\[3.5mm]
 F_2^{\mbox{\tiny R}}(s,t) & = & - \, \frac{\sqrt{2}}{6} \, \frac{F_V\,G_V}{F^3}  \, \left[
 \, \frac{B^{\mbox{\tiny R}}(t,u,m_K^2,m_K^2)}{M_{\rho}^{2}-s}  \,  +
 \, \frac{A^{\mbox{\tiny R}}(Q^2,t,u,m_K^2,m_K^2,m_{\pi}^2)}{M_{K^{*}}^{2}-t} \,\right] ,  \,  \\[4mm]
F_2^{\mbox{\tiny RR}}(s,t) & = & \frac{2}{3} \, \frac{F_A G_V}{F^3} \, \frac{Q^2}{M_{a_1}^2-Q^2} \,
\, \left[ \, \frac{B^{\mbox{\tiny RR}}(Q^2,t,u,s,m_K^2,m_K^2,m_{\pi}^2) }{M_{\rho}^2-s} \,
\right. \nonumber \\
& &  \qquad \qquad \qquad \qquad \qquad\left.
+ \, \frac{A^{\mbox{\tiny RR}}(Q^2,t,u,m_K^2,m_K^2,m_{\pi}^2) }{M_{K^*}^2-t} \, \right] \, \, . \nonumber
\end{eqnarray}
The $F_3$ form factor arises from the chiral anomaly and the non-anomalous
odd-intrinsic-parity amplitude. We obtain~:
\begin{eqnarray}
\label{eq:f31}
 F_3^{\chi} & = & - \frac{N_C \, \sqrt{2}}{12 \, \pi^2 \, F^3} \, , \nonumber \\[3.5mm]
 F_3^{\mbox{\tiny R}}(s,t) & = & - \frac{4 \, G_V}{M_V \, F^3} \, \left[ \,
C^{\mbox{\tiny R}}(Q^2,s,m_K^2,m_K^2,m_{\pi}^2) \,
\left( \sin^2 \theta_V \frac{1+ \sqrt{2}  \cot \theta_V}{M_{\omega}^2-s}
\right.\right. \nonumber \\
& & \qquad \qquad \; \; \left. + \, \cos^2 \theta_V
\frac{1- \sqrt{2} \tan \theta_V }{M_{\phi}^2-s} \right)
\, + \, \frac{C^{\mbox{\tiny R}}(Q^2,t,m_K^2,m_\pi^2,m_K^2)}{M_{K^*}^2-t}
\nonumber \\
& & \qquad \qquad \; \; \left. \, - \, \frac{2 \, F_V}{G_V}
\, \frac{D^{\mbox{\tiny R}}(Q^2,s,t)}{M_{\rho}^2-Q^2} \right] \, , \\[3.5mm]
F_3^{\mbox{\tiny RR}}(s,t) & = & 4 \sqrt{2} \frac{F_V \, G_V}{F^3} \,  \frac{1}{M_{\rho}^2 - Q^2} \,
\left[ C^{\mbox{\tiny RR}}(Q^2,s,m_{\pi}^2) \,
\left( \sin^2 \theta_V \frac{1+ \sqrt{2} \cot \theta_V}{M_{\omega}^2-s} \right. \right.
\nonumber \\
&& \qquad \qquad \qquad \qquad \qquad \; \left. \left. + \, \cos^2 \theta_V
\frac{1- \sqrt{2} \tan \theta_V }{M_{\phi}^2 -s} \right)
\, + \, \frac{C^{\mbox{\tiny RR}}(Q^2,t,m_K^2)}{M_{K^*}^2-t} \right] \, ,
\nonumber
\end{eqnarray}
where $C^{\mbox{\tiny R}}$, $D^{\mbox{\tiny R}}$ and $C^{\mbox{\tiny RR}}$
are defined in Appendix~\ref{ap:1}, and $\theta_V$ is the mixing angle
between the octet and singlet vector states $\omega_8$ and $\omega_0$
that defines the mass eigenstates $\omega(782)$ and $\phi(1020)$~:
\begin{equation}
\left(  \begin{array}{c}
  \phi \\
  \omega
 \end{array} \right)\, = \,  \left(
\begin{array}{cc}
\cos \theta_V & - \sin \theta_V \\
\sin \theta_V &  \cos \theta_V
\end{array}  \right) \;  \left(
\begin{array}{c}
 \omega_8 \\
\omega_0
\end{array} \right)\, .
\end{equation}
For numerical evaluations we will assume ideal mixing,  i.e.\ $\theta_V
= \tan^{-1}(1/\sqrt{2})$. In this case the contribution of the $\phi(1020)$
meson to $F_3$ vanishes.
\par
Finally, though we have not dwelled on specific contributions to the $F_4$
form factor, we quote for completeness the result obtained from our
Lagrangian. Its structure is driven by the pion pole~:
\begin{eqnarray}
\label{eq:f41}
 F_4 & = & F_4^{\chi} + F_4^{\mbox{\tiny R}} \, , \nonumber \\[3mm]
 F_4^{\chi}(s,t) & = & \frac{1}{\sqrt{2} \, F} \frac{m_{\pi}^2}{m_{\pi}^2 - Q^2} \,
\left( 1+ \frac{m_K^2 - u}{Q^2} \right) \, , \nonumber \\[3mm]
 F_4^{\mbox{\tiny R}}(s,t) & = & \frac{G_V^2}{\sqrt{2} \, F^3}
\frac{m_{\pi}^2}{Q^2 (m_{\pi}^2 - Q^2)} \, \left[ \frac{s(t-u)}{M_{\rho}^2 -s}
+ \frac{t(s-u) - (m_K^2 - m_\pi^2)(Q^2-m_K^2)}{M_{K^*}^2-t} \right] \, .
\end{eqnarray}

\subsection{Form factors in $\tau^-  \rightarrow  K^- \,K^0 \, \pi^0 \, \nu_{\tau}$}

The diagrams contributing to the $\tau^-  \rightarrow  K^- \,K^0 \, \pi^0
\, \nu_{\tau}$ decay amplitude are also those in Fig.~\ref{fig:feynman},
hence once again we can write $F_i \, =\, F_i^{\chi} \, + \,
F_i^{\mbox{\tiny R}} \, + \, F_i^{\mbox{\tiny RR}} \, + \, \dots$. However,
the structure of the form factors for this process does not show the
symmetry observed in $\tau \rightarrow K \overline{K} \pi \nu_{\tau}$. We
find~:
\begin{eqnarray}
\label{eq:f12}
F_1^{\chi}  & = & -\frac{1}{F} \, , \nonumber \\[3mm]
F_1^{\mbox{\tiny R}}(s,t) & = & - \frac{1}{6} \frac{F_V G_V}{F^3} \, \left[
\, \frac{B^{\mbox{\tiny R}}(s,u,m_K^2,m_{\pi}^2)}{M_{K^{*}}^{2}-t} \, +
\, 2 \; \frac{A^{\mbox{\tiny R}}(Q^2,s,u,m_K^2,m_\pi^2,m_K^2)}{M_\rho^{2}-s} \,
\right. \nonumber \\
& & \left. \qquad\qquad\qquad +
\, \frac{A^{\mbox{\tiny R}}(Q^2,u,s,m_\pi^2,m_K^2,m_K^2)}{M_{K^{*}}^{2}-u} \, \right]
\, , \nonumber \\ [3.5mm]
F_1^{\mbox{\tiny RR}}(s,t) & = &  \frac{\sqrt{2}}{3} \frac{F_A G_V}{F^3}
\frac{Q^2}{M_{a_1}^2-Q^2} \, \left[
\, \frac{B^{\mbox{\tiny RR}}(Q^2,s,u,t,m_K^2,m_{\pi}^2,m_K^2)}{M_{K^{*}}^{2}-t}
\right. \nonumber \\
& & \qquad\qquad\qquad\qquad\qquad
 + \, 2 \, \frac{A^{\mbox{\tiny RR}}(Q^2,s,u,m_K^2,m_\pi^2,m_K^2)}{M_\rho^{2}-s} \,
\nonumber \\
& & \left.\qquad\qquad\qquad\qquad\qquad
 + \, \frac{A^{\mbox{\tiny RR}}(Q^2,u,s,m_\pi^2,m_K^2,m_K^2)}{M_{K^{*}}^{2}-u} \, \right]
\, ,
\end{eqnarray}
\begin{eqnarray}
\label{eq:f22}
 F_2^{\chi} & = & 0 \, , \nonumber \\ [3mm]
F_2^{\mbox{\tiny R}}(s,t) & = & - \frac{1}{6} \frac{F_V G_V}{F^3} \, \left[
\, \frac{A^{\mbox{\tiny R}}(Q^2,t,u,m_K^2,m_K^2,m_{\pi}^2)}{M_{K^{*}}^{2}-t} \, +
\, 2 \; \frac{B^{\mbox{\tiny R}}(t,u,m_K^2,m_K^2)}{M_\rho^{2}-s} \,
\right. \nonumber \\
& & \left. \qquad\qquad\qquad -
\, \frac{A^{\mbox{\tiny R}}(Q^2,u,t,m_K^2,m_K^2,m_\pi^2)}{M_{K^{*}}^{2}-u} \, \right]
\, , \nonumber \\ [3.5mm]
F_2^{\mbox{\tiny RR}}(s,t) & = &  \frac{\sqrt{2}}{3} \frac{F_A G_V}{F^3}
\frac{Q^2}{M_{a_1}^2-Q^2} \, \left[
\, \frac{A^{\mbox{\tiny RR}}(Q^2,t,u,m_K^2,m_K^2,m_{\pi}^2)}{M_{K^{*}}^{2}-t}
\right. \nonumber \\
& & \qquad\qquad\qquad\qquad\qquad
+ \, 2 \; \frac{B^{\mbox{\tiny RR}}(Q^2,t,u,s,m_K^2,m_K^2,m_\pi^2)}{M_\rho^{2}-s} \,
\nonumber \\
& & \left. \qquad\qquad\qquad\qquad\qquad
- \, \frac{A^{\mbox{\tiny RR}}(Q^2,u,t,m_K^2,m_K^2,m_\pi^2)}{M_{K^{*}}^{2}-u} \, \right] \, .
\end{eqnarray}
\par
The form factor driven by the vector current is given by~:
\begin{eqnarray}
\label{eq:f32}
 F_3^{\chi} & = & 0  \, \nonumber \\ [3mm]
F_3^{\mbox{\tiny R}}(s,t) & = & \frac{2 \sqrt{2} \, G_V}{M_V \, F^3}
\!\!\left[ \frac{C^{\mbox{\tiny R}}(Q^2,t,m_K^2,m_\pi^2,m_K^2)}{M_{K^*}^2-t} -
\frac{C^{\mbox{\tiny R}}(Q^2,u,m_K^2,m_\pi^2,m_K^2)}{M_{K^*}^2-u}
 - \frac{2 F_V}{G_V}
\frac{E^{\mbox{\tiny R}}(t,u)}{M_{\rho}^2-Q^2} \right] \, , \nonumber \\ [3mm]
F_3^{\mbox{\tiny RR}}(s,t) & = & - 4 \frac{F_V G_V}{F^3} \frac{1}{M_{\rho}^2-Q^2}
\left[ \frac{C^{\mbox{\tiny RR}}(Q^2,t,m_K^2)}{M_{K^*}^2-t} -
\frac{C^{\mbox{\tiny RR}}(Q^2,u,m_K^2)}{M_{K^*}^2-u} \right] \, ,
\end{eqnarray}
with $E^{\mbox{\tiny R}}$ defined in Appendix~\ref{ap:1}.
\par
Finally for the pseudoscalar form factor we have~:
\begin{eqnarray}
\label{eq:f42}
 F_4^{\chi}(s,t) & = & \frac{1}{2 \, F} \frac{m_{\pi}^2 \, (t-u)}{Q^2 ( m_{\pi}^2 - Q^2 )} \, , \nonumber \\ [3mm]
F_4^{\mbox{\tiny R}}(s,t) & = & \frac{1}{2} \frac{G_V^2}{F^3} \frac{m_{\pi}^2}{Q^2 (m_{\pi}^2 - Q^2 )}
\left[ \frac{t(s-u)-(m_K^2-m_{\pi}^2)(Q^2-m_K^2)}{M_{K^*}^2-t}
+ \frac{2 \, s (t-u)}{M_{\rho}^2-s} \right. \nonumber \\ [3mm]
& & \qquad \qquad \qquad \qquad \; \; \; \,\left. - \frac{u(s-t)-(m_{K}^2-m_{\pi}^2)(Q^2-m_K^2)}{M_{K^*}^2-u}
\right] \, .
\end{eqnarray}

\subsection{Features of the form factors}

Several remarks are needed in order to understand our previous results for
the form factors related with the vector and axial-vector QCD currents
analysed above~:
\begin{itemize}
 \item[1/] Our evaluation corresponds to the tree level diagrams in Fig.~\ref{fig:feynman} that
arise from the $N_C \rightarrow \infty$ limit of QCD. Hence the masses of the resonances
would be reduced to $M_V=M_{\rho}=M_{\omega}=M_{K^*}=M_{\phi}$ and $M_A=M_{a_1}$ as they appear in the resonance Lagrangian (\ref{eq:lag0}), i.e.
the masses of the nonet of vector and axial-vector resonances in the chiral and large-$N_C$ limit.
However it is easy to introduce NLO corrections in the $1/N_C$ and chiral expansions on the masses by including
the {\em physical} ones~: $M_{\rho}$, $M_{K^*}$, $M_{\omega}$, $M_{\phi}$ and $M_{a_1}$ for
the $\rho(770)$, $K^*(892)$, $\omega(782)$, $\phi(1020)$ and $a_1(1260)$ states, respectively,
as we have done in the expressions of the form factors.
In this setting
resonances also have zero width, which represents a drawback if we intend to
analyse the phenomenology of the processes~: Due to the high mass of the tau
lepton, resonances do indeed resonate producing divergences if their width
is ignored. Hence we will include energy-dependent widths for the
$\rho(770)$, $a_1(1260)$ and $K^*(892)$ resonances, that are rather wide,
and a constant width for the $\omega(782)$. This
issue is discussed in Ref.~\cite{shortp}.
\par
In summary, to account for the inclusion of NLO corrections we perform the
substitutions~:
\begin{equation}
 \frac{1}{M_R^2-q^2} \; \; \longrightarrow \;\; \frac{1}{M_{phys}^2-q^2- \, i \, M_{phys} \, \Gamma_{phys}(q^2)} \; ,
\end{equation}
where $R=V,A$, and the subindex {\em phys} on the right hand side stands for
the corresponding {\em physical} state depending on the relevant Feynman
diagram.

\item[2/] If we compare our results with those of Ref.~\cite{KS5}, evaluated
within the KS model, we notice that the structure of our form factors is
fairly different and much more intricate. This is due to the fact that the
KS model, i.e.\ a model resulting from combinations of {\em ad hoc}
products of Breit-Wigner functions, does not meet higher order chiral constraints enforced in
our approach.
\item[3/] As commented above the pseudoscalar form factors $F_4$ vanishes in the chiral limit. Indeed
the results of Eqs.~(\ref{eq:f41}, \ref{eq:f42}) show that they are
proportional to $m_{\pi}^2$, which is tiny compared with any other scale in
the amplitudes. Hence the contribution of $F_4$ to the structure of the
spectra is actually marginal.
\end{itemize}

\section{QCD constraints and determination of resonance coupling constants}
\label{sect:4}

\hspace*{0.5cm} Our results for the form factors $F_i$ depend on several
combinations of the coupling constants in our Lagrangian ${\cal L}_{R \chi
T}$ in Eq.~(\ref{eq:ourtheory}), most of which are in principle unknown
parameters. Now, if our theory offers an adequate effective description of
QCD at hadron energies, the underlying theory of the strong interactions
should give information on those constants. Unfortunately the
determination of the effective parameters from first principles is still an
open problem in hadron physics.
\par
A fruitful procedure when working with resonance Lagrangians has been to
assume that the resonance region, even when one does not include the full
phenomenological spectrum, provides a bridge between the chiral and
perturbative regimes \cite{Ecker:1989yg}. The chiral constraints supply
information on the structure of the interaction but do not provide any hint
on the coupling constants of the Lagrangian. Indeed, as in any effective
theory \cite{Georgi:1991ch}, the couplings encode information from high
energy dynamics. Our
procedure amounts to match the high energy behaviour of Green functions (or
related form factors) evaluated within the resonance theory with the
asymptotic results of perturbative QCD. This strategy has proven to be
phenomenologically sound
\cite{Moussallam:1997xx,Knecht:2001xc,RuizFemenia:2003hm,Cirigliano:2004ue,Cirigliano:2005xn,Mateu:2007tr,
Ecker:1989yg,Amoros:2001gf},
and it will be applied here in order to obtain information on the unknown
couplings.
\par
Two-point Green functions of vector and axial-vector currents
$\Pi_{V,A}(q^2)$ were studied within perturbative QCD in
Ref.~\cite{Floratos:1978jb}, where it was shown that both spectral functions
go to a constant value at infinite transfer of momenta~:
\begin{equation}
 \Im m \, \Pi_{V,A}(q^2) \, \mapright{}{\; \; \; q^2 \rightarrow \infty \; \; \; } \, \, \frac{N_C}{12 \, \pi}
\, .
\end{equation}
By local duality interpretation  the imaginary part of the quark loop
can be understood as the sum of infinite positive contributions of
intermediate hadron states. Now, if the infinite sum is going to behave like
a constant at $q^2 \rightarrow \infty$, it is heuristically sound to expect
that each one of the infinite contributions vanishes in that limit.
This deduction stems from the fact that vector and axial-vector form factors
should behave smoothly at high $q^2$, a result previously put forward from
parton dynamics in Ref.~\cite{Brodsky}. Accordingly in the $N_C \rightarrow
\infty$ limit this result applies to our form factors evaluated at tree
level in our framework.
\par
Other hints involving short-distance dynamics may also be considered. The
analyses of three-point Green functions of QCD currents have become a
useful procedure to determine coupling constants  of the intermediate
energy (resonance) framework
\cite{Moussallam:1997xx,Knecht:2001xc,RuizFemenia:2003hm,Cirigliano:2004ue,Cirigliano:2005xn}.
The idea is to use those functions (order parameters of the chiral symmetry
breaking), evaluate them within the resonance framework and match this
result with the leading term in the Operator Product Expansion (OPE) of the
Green function.
\par
In the following we collect the information provided by these hints on our
coupling constants, attaching always to the $N_C \rightarrow \infty$ case
\cite{Pich:2002xy}
(approximated with only one nonet of vector and axial-vector resonances)~:
\begin{itemize}
 \item[i)] By demanding that the two-pion vector form factor vanishes at high $q^2$ one obtains the
condition $F_V \, G_V = F^2$ involving the couplings in Eq.~(\ref{eq:lag1}) \cite{Ecker:1989yg}.
\item[ii)] The first Weinberg sum rule \cite{Weinberg:1967kj} leads to $F_V^2 - F_A^2 = F^2$, and
the second Weinberg sum rule gives $F_V^2 \, M_V^2 \, = \, F_A^2 \, M_A^2$
\cite{Ecker:1988te}.
\item[iii)] The analysis of the VAP Green function \cite{Cirigliano:2004ue} gives for the combinations
of couplings defined in Eq.~(\ref{eq:lambdias}) the following results~:
\begin{eqnarray}
\label{eq:lambres}
 \lambda' & = & \frac{F^2}{2 \, \sqrt{2} \, F_A \, G_V} \; = \; 
\frac{M_A}{2 \, \sqrt{2} \, M_V} \,, \nonumber \\[3.5mm]
\lambda'' & = & \frac{2 \, G_V \, - F_V}{2 \, \sqrt{2} \, F_A} \; = \; \frac{M_A^2 - 2 M_V^2}{2 \, \sqrt{2} \, M_V \, M_A} \, , \nonumber \\[3.5mm]
4 \, \lambda_0 & = & \lambda' + \lambda'' \; ,
\end{eqnarray}
where, in the two first relations, the second equalities come from using relations i) and ii) above.
Here $M_V$ and $M_A$ are the masses appearing in the resonance Lagrangian (\ref{eq:lag0}).
Contrarily to what happens in the vector case where $M_V$
is well approximated by the $\rho(770)$ mass,  in
Ref.~\cite{Mateu:2007tr} it was obtained $M_A = 998 (49) \, \mbox{MeV}$,
hence $M_A$ differs appreciably from the presently accepted value of
$M_{a_1}$. It is worth to notice that the two first relations in
Eq.~(\ref{eq:lambres}) can also be obtained from the requirement that
the $J=1$ axial spectral function in $\tau \rightarrow 3 \pi \nu_{\tau}$
vanishes for large momentum transfer \cite{GomezDumm:2003ku}.
\item[iv)] Both vector form factors contributing to the final states $K \overline{K} \pi^-$ and
$K^- K^0 \pi^0$ in tau decays, when integrated over the available phase space, should also vanish at high $Q^2$. Let us consider $H_{\mu \nu}^3(s,t,Q^2) \equiv T_{\mu}^3 T_{\nu}^{3 \, *}$, where $T_{\mu}^3$ can be
inferred from Eq.~(\ref{eq:t3}). Then we define $\Pi_V(Q^2)$ by~:
\begin{equation}
 \int \, \mbox{d}\Pi_3 \, H_{\mu \nu}^3(s,t,Q^2) \, = \, \left( Q^2 g_{\mu \nu} \, - \,
Q_{\mu} Q_{\nu} \right) \, \Pi_V(Q^2) \, ,
\end{equation}
where
\begin{eqnarray}
 \int \mathrm{d}\Pi_3 \, & = & \, \int \frac{d^3 p_1}{2 E_1} \frac{d^3 p_2}{2 E_2} \frac{d^3 p_3}{2 E_3}
\delta^4\left( Q-p_1-p_2-p_3\right)  \delta \left( s-(Q-p_3)^2 \right) \delta \left( t-(Q-p_2)^2 \right) \,
\nonumber \\
\, & = &  \, \frac{\pi^2}{4 \,Q^2} \, \int ds \, dt \; .
\end{eqnarray}
Hence we find that
\begin{equation}
 \Pi_V(Q^2) \, = \, \frac{\pi^2}{12 \, Q^4} \, \int \, \mathrm{ds \, dt} \, g^{\mu \nu} \,
H_{\mu \nu}^3(s,t,Q^2) \, ,
\end{equation}
where the limits of integration are those of Eq.~(\ref{eq:phaspace1}, \ref{eq:phaspace2}), should vanish
at $Q^2 \rightarrow \infty$. This constraint determines several relations on the couplings that
appear in the $F_3$ form factor, namely~:
\begin{eqnarray}
\label{eq:1c}
 c_1 \, - \, c_2 \, + \, c_5 \, & = & 0 \, , \\
\label{eq:2c}
c_1 \, - \, c_2 \, - \, c_5 \, + \, 2 c_6 \, & = & - \, \frac{ \,N_C}{96 \, \pi^2} \,
\frac{F_V \, M_V}{\sqrt{2} \, F^2} \, , \\
\label{eq:3c}
d_3 & = & - \frac{N_C}{192 \, \pi^2} \, \frac{M_V^2}{F^2} \, , \\
\label{eq:4c}
g_1 \, + \, 2 g_2 \, - g_3 & = & 0 \, , \\
\label{eq:5c}
g_2 & = & \frac{N_C}{192 \,\sqrt{2} \, \pi^2} \, \frac{M_V}{F_V} \, .
\end{eqnarray}
If these conditions are satisfied, $\Pi_V(Q^2)$ vanishes at high
transfer of momenta for both $K \overline{K} \pi^-$ and $K^- K^0 \pi^0$
final states. We notice that the result in Eq.~(\ref{eq:1c}) is in
agreement with the corresponding relation in Ref.~\cite{RuizFemenia:2003hm},
while Eqs.~(\ref{eq:2c}) and (\ref{eq:3c}) do not agree with the results in
that work. In this regard we point out that the relations in
Ref.~\cite{RuizFemenia:2003hm}, though they satisfy the leading matching to the
OPE expansion of the $\langle VVP \rangle$ Green function with the inclusion
of one multiplet of vector mesons, do not reproduce the right asymptotic
behaviour of related form factors. Indeed it has been shown
\cite{Knecht:2001xc,Mateu:2007tr} that two multiplets of vector resonances
are needed to satisfy both constraints. Hence we will attach to our results
above, which we consider more reliable~\footnote{One of the form factors
derived from the $\langle VVP \rangle$ Green function is ${\cal F}_{\pi
\gamma^* \gamma}(q^2)$, that does not vanish at high $q^2$ with the set of
relations in Ref.~\cite{RuizFemenia:2003hm}. With our conditions in
Eqs.~(\ref{eq:2c},\ref{eq:3c}) the asymptotic constraint on the form factor
can be satisfied if the large-$N_C$ masses, $M_A$ and $M_V$, fulfill the
relation $2 M_A^2 = 3 M_V^2$. It is interesting to notice the significant
agreement with the numerical values for these masses mentioned above.}.
\item[v)] An analogous exercise to the one in iv) can be carried out for the axial-vector form
factors $F_1$ and $F_2$. We have performed such an analysis and, using the relations in i) and ii) above,
it gives us back the results provided in
Eq.~(\ref{eq:lambres}) for $\lambda'$ and $\lambda''$. Hence both procedures
give a consistent set of relations.
\end{itemize}
After imposing the above constraints, let us analyse which coupling
combinations appearing in our expressions for the form factors are still
unknown. We intend to write all the information on the couplings in terms
of $F$, $M_V$ and $M_A$. From the relations involving $F_V$, $F_A$ and $G_V$
we obtain~:
\begin{eqnarray} \label{eq:fvfagv}
 \frac{F_V^2}{F^2} & = & \frac{M_A^2}{M_A^2-M_V^2} \, , \nonumber \\
\frac{F_A^2}{F^2} & = & \frac{M_V^2}{M_A^2-M_V^2} \, , \nonumber \\
\frac{G_V^2}{F^2} & = & 1 \, - \, \frac{M_V^2}{M_A^2} \, .
\end{eqnarray}
Moreover we know that $F_V$ and $G_V$ have the same sign, and we will assume
that it is also the sign of $F_A$. Together with the relations in
Eq.~(\ref{eq:lambres}) this determines completely the axial-vector form
factors $F_{1,2}$. Now from Eqs.~(\ref{eq:1c}-\ref{eq:5c}) one can fix all
the dominant pieces in the vector form factor $F_3$, i.e. those pieces that
involve factors of the kinematical variables $s$, $t$ or $Q^2$. The unknown
terms, that carry factors of $m_{\pi}^2$ or $m_K^2$, are expected to be less
relevant. They are given by the combinations of couplings~:
 $c_1 + c_2 + 8 \, c_3 - c_5$,  $d_1 + 8 \, d_2$,  $c_4$ , $g_4$ and  $g_5$.
However small they may be, we will not neglect these contributions, and we
will proceed as follows. Results in Ref.~\cite{RuizFemenia:2003hm} determine
the first and the second coupling combinations. As commented above the
constraints in that reference do not agree with those we have obtained by
requiring that the vector form factor vanishes at high $Q^2$. However, they
provide us an estimate to evaluate terms that, we recall, are suppressed by
pseudoscalar masses. In this way, from a phenomenological analysis of
$\omega \rightarrow \pi^+ \pi^- \pi^0$ (see Appendix~\ref{ap:3}) it is
possible to determine the combination $2 \, g_4 + g_5$. Finally in order to
evaluate $c_4$ and $g_4$ we will combine the recent analysis of $\sigma
\left( e^+ e^- \rightarrow K K \pi \right)$ by BABAR \cite{Aubert:2007ym}
with the information from the $\tau \rightarrow K K \pi \nu_{\tau}$ width.

\subsection{Determination of $c_4$ and $g_4$}
\label{sect:41}

The separation of isoscalar and isovector components of the $e^+ e^- \rightarrow K K \pi$ amplitudes,
carried out by BABAR \cite{Aubert:2007ym}, provides us with an additional tool for the estimation
of the coupling constant $c_4$ that appears in the hadronization of the vector current.
Indeed, using $SU(2)_I$ symmetry alone one can relate the isovector contribution to
$\sigma \left( e^+ e^- \rightarrow K^- K^0 \pi^+ \right)$ with the vector contribution to
$\Gamma \left( \tau^- \rightarrow K^0 K^- \pi^0 \nu_{\tau} \right)$ through the relation~:
\begin{equation} \label{eq:CVC}
  \frac{\mathrm{d}}{\mathrm{d} \, Q^2} \, \Gamma \left( \tau^- \rightarrow K^0 K^- \pi^0 \nu_{\tau} \right)
\Bigg|_{F_3} \, = \, f(Q^2) \; \sigma_{I=1} \left( e^+ e^- \rightarrow K^- K^0 \pi^+ \right)
\; ,
\end{equation}
where $f(Q^2)$ is given in Appendix~\ref{ap:4}. In this Appendix we also
discuss other relations similar to Eq.~(\ref{eq:CVC}) that have been used in
the literature and we point out the assumptions on which they rely.
\par
\begin{figure}[!t]
\begin{center}
\vspace*{0.2cm}
\includegraphics[scale=0.45,angle=-90]{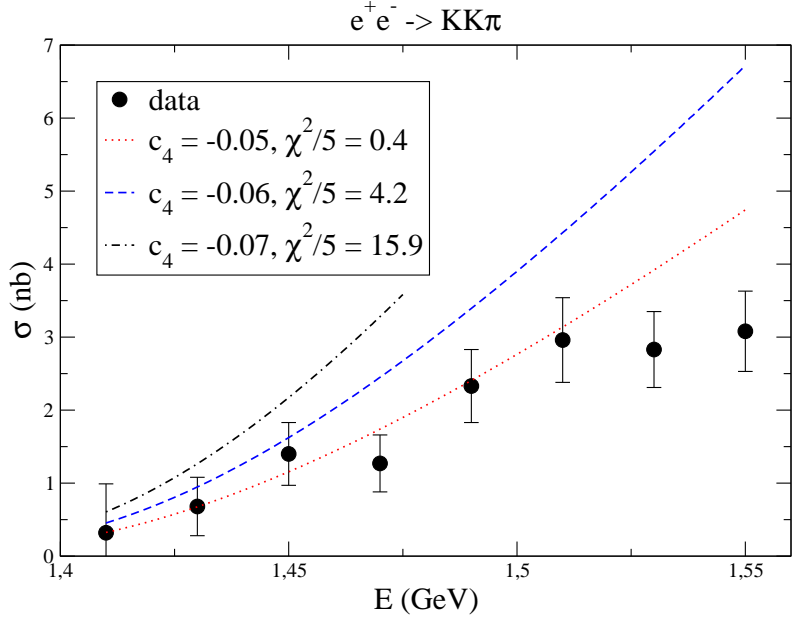}
\caption[]{\label{fig:ee} \small{Comparison of the experimental data
\cite{Aubert:2007ym} with the theoretical prediction for the cross-section
of the isovector component of $e^+ e^-  \rightarrow K^*(892) K \rightarrow
K_S K^{\pm} \pi^{\mp}$ process, for different values of the $c_4$ coupling.
The $\chi^2$ values are associated to the first 6 data points only.
}}
\end{center}
\end{figure}

Hence we could use the isovector contribution to the cross-section for the
process $e^+ e^- \rightarrow K_S K^{\pm} \pi^{\mp}$ determined by BABAR and
Eq.~(\ref{eq:CVC}) to fit the $c_4$ coupling that is the only still
undetermined constant in that process. However we have to take into
account that our description for the hadronization of the vector current in
the tau decay channel does not, necessarily, provide an adequate description
of the cross-section. Indeed the complete different kinematics of both
observables suppresses the high-energy behaviour of the bounded tau decay
spectrum,  while this suppression does not occur in the cross-section.
Accordingly, our description of the latter away from the energy
threshold can be much poorer. As can be seen in Fig.~\ref{fig:ee} there is
a clear structure in the experimental points of the cross-section that 
is not provided by our description.
\par
Taking into account the input parameters quoted in Eq.~(\ref{eq:set1}) we
obtain~: $ c_4 \, = \, -0.047 \pm 0.002 $. The fit has been carried out for
the first 6 bins (up to $E_{cm} \sim 1.52 \, \mbox{GeV}$). This result
corresponds to $\chi^2 /dof = 0.3$ and the displayed error comes only from
the fit.
\par
We take into consideration now the measured branching ratios for the $K K
\pi$ channels of Table~\ref{tab:52} in order to extract information both
from $c_4$ and $g_4$. We notice that it is not possible to reconcile a
prediction of the branching ratios of $\tau \rightarrow K \overline{K} \pi
\nu_{\tau}$ and $\tau \rightarrow K^- K^0 \pi^0 \nu_{\tau}$ in spite of the
noticeable size of the errors shown in the Table~\ref{tab:52}. Considering
that the second process was measured long ago and that the $\tau^-
\rightarrow K^+ K^- \pi^- \nu_{\tau}$ decay has been focused by both CLEO
III and BABAR we intend to fit the branching ratio of the latter. For
the parameter values~:
\begin{eqnarray} \label{eq:c4g4I}
 c_4 & = & -0.07 \pm 0.01 \, , \nonumber \\
 g_4 & = & -0.72 \pm 0.20 \, ,
\end{eqnarray}
we find a good agreement with the measured widths $\Gamma(\tau^- \rightarrow
K^+ K^- \pi^- \nu_{\tau})$ and $\Gamma(\tau \rightarrow K^- K^0 \pi^0 \nu_{\tau})$
within errors (see Table~\ref{tab:52}). Notice that the value of $|c_4|$ is
larger than that obtained from the fit to the $e^+ e^- \rightarrow K_S
K^{\pm} \pi^{\mp}$ data explained above. In Fig.~\ref{fig:ee} we show the
first 8 bins in the isovector component of $e^+ e^- \rightarrow K_S K^{\pm}
\pi^{\mp}$ and the theoretical curves for different values of the $c_4$
coupling. As our preferred result we choose the larger value of $c_4$ in
Eq.~(\ref{eq:c4g4I}), since it provides a better agreement with the present
measurement of $\Gamma(\tau^- \rightarrow K^- K^0 \pi^0 \nu_{\tau})$.
Actually, one can expect a large incertitude in the splitting of isospin
amplitudes in the $e^+ e^- \rightarrow K_S K^{\pm} \pi^{\mp}$ cross-section
(see Appendix~\ref{ap:4}). Taking into account this systematic error, it
could be likely that the theoretical curve with $c_4 = -0.07$ falls within
the error bars for the first data points.

\section{Phenomenology of $\tau \rightarrow K K \pi \nu_{\tau}$~: Results and their analysis}
\label{sect:5}

Asymmetric B-factories span an ambitious $\tau$ programme that includes the
determination of the hadron structure of semileptonic $\tau$ decays
such as the $K K \pi$ channel. As commented in the Introduction the
latest study of $\tau^- \rightarrow K^+ K^- \pi^- \nu_{\tau}$ by the CLEO
III Collaboration \cite{Liu:2002mn} showed a disagreement between the KS
model, included in TAUOLA, and the data. Experiments with higher statistics
such as BABAR and Belle should clarify the theoretical settings.
\par
For the numerics in this Section we use the values in Appendix~\ref{ap:input}.
At present no spectra for these channels is available and the
determinations of the widths are collected in Table~\ref{tab:52}.
\begin{table}
\begin{center}
\begin{tabular}{|c|c|c|c|}
\hline
&&& \\[-2.5mm]
Source & $\Gamma ( \tau^-\to K^+ K^- \pi^- \nu_\tau) \,$
& $ \Gamma ( \tau^-\to K^0 \overline{K}^0 \pi^- \nu_\tau) \,$
 & $\Gamma ( \tau^-\to K^- K^0 \pi^0 \nu_\tau) \,$  \\ [1.5mm]
\hline
&&&\\ [-2.5mm]
 PDG  \cite{PDG2008} & $3.103\,(136)$& $3.465 \, (770) $& $3.262 \, (521)$ \\ [1.9mm]
BABAR \cite{:2007mh} &$3.049 \, (85)$ & &  \\
[1.9mm]
CLEO III \cite{Liu:2002mn}&$3.511 \, (245)$ & &  \\ [1.9mm]
Belle \cite{:2008sg} & $3.465 \, (136)$ && \\ [1.9mm]
Our prediction & $3.4^{+0.5}_{-0.2}$ & $3.4^{+0.5}_{-0.2}$ & $2.5^{+0.3}_{-0.2}$ \\ [2mm]
\hline
\end{tabular}
\caption{\small{Comparison of the measurements of partial widths (in units
of $10^{-15} \, \mbox{GeV}$) with our predictions for the set of values in
Eq.~(\ref{eq:c4g4I}). For earlier references see \cite{PDG2008}.}}
\label{tab:52}
\end{center}
\end{table}
We also notice that there is a discrepancy between the BABAR measurement of
$\Gamma(\tau^- \rightarrow K^+ K^- \pi^- \nu_{\tau})$ and the results by
CLEO and Belle. Within $SU(2)$ isospin symmetry it is found that
$\Gamma(\tau^- \rightarrow K^+ K^- \pi^- \nu_{\tau}) = \Gamma(\tau^-
\rightarrow K^0 \overline{K}^0 \pi^- \nu_{\tau})$, which is well reflected
by the values in Table~\ref{tab:52} within errors. Moreover, as commented
above, the PDG data \cite{PDG2008} indicate that $\Gamma(\tau^-
\rightarrow K^- K^0 \pi^0 \nu_{\tau})$ should be similar to $\Gamma(\tau^-
\rightarrow K \overline{K} \pi \nu_{\tau})$. It would be
important to obtain a more accurate determination of the $\tau^-\rightarrow
K^- K^0 \pi^0 \nu_{\tau}$ width (the measurements quoted by the PDG are
rather old) in the near future.
\par
In our analyses we include the lightest resonances in both the vector and
axial-vector channels, namely $\rho(775)$, $K^*(892)$ and $a_1(1260)$. It is
clear that, as it happens in the $\tau \rightarrow \pi  \pi \pi \nu_{\tau}$
channel (see Ref.~\cite{shortp}), a much lesser role, though noticeable, can
be played by higher excitations on the vector channel. As experimentally
only the branching ratios are available for the $K K \pi$ channel we think
that the refinement of including higher mass resonances should be taken into
account in a later stage, when the experimental situation improves.
\par
\begin{figure}[!t]
\begin{center}
\vspace*{0.2cm}
\includegraphics[scale=0.45,angle=-90]{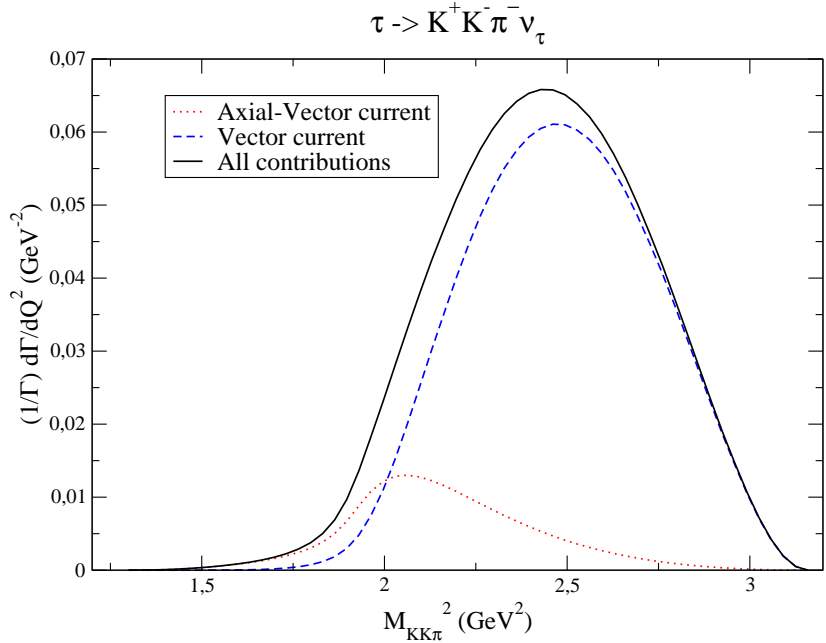}
\caption[]{\label{fig:53} \small{Normalized $M_{K K \pi}^2$-spectra for
$\tau^- \rightarrow K^+ K^- \pi^- \nu_{\tau}$. Notice the dominance of the axial-vector current
at very low values of $Q^2$.}}
\end{center}
\end{figure}
\begin{figure}[!ht]
\begin{center}
\vspace*{0.2cm}
\includegraphics[scale=0.45,angle=-90]{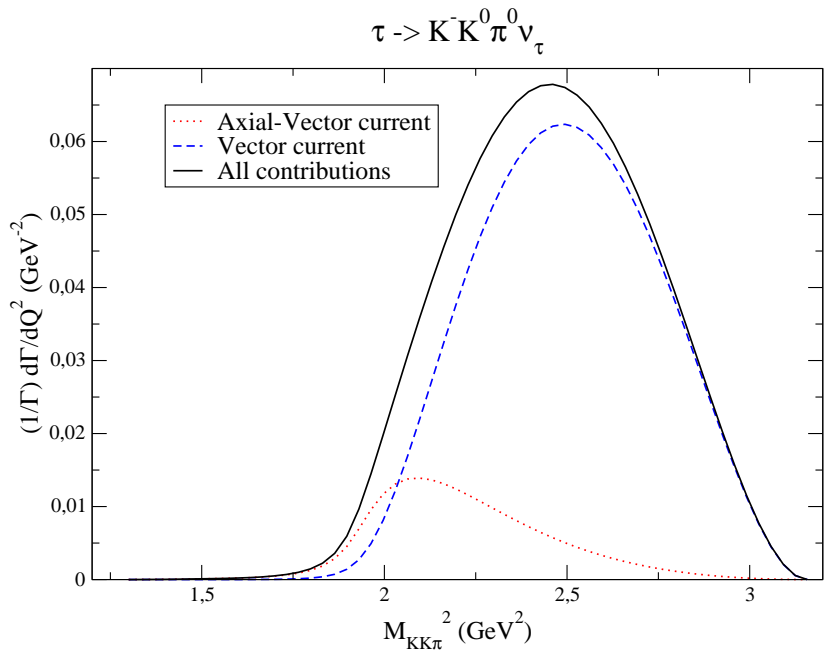}
\caption[]{\label{fig:54} \small{Normalized $M_{K K \pi}^2$-spectra for
$\tau^- \rightarrow K^-K^0\pi^0\nu_{\tau}$. Notice the dominance of the axial-vector current
at very low values of $Q^2$.}}
\end{center}
\end{figure}

In Figs.~\ref{fig:53} and \ref{fig:54} we show our predictions for the
normalized $M_{K K \pi}^2-$spectrum of the $\tau^- \rightarrow K^+ K^- \pi^-
\nu_{\tau}$ and $\tau^- \rightarrow K^-K^0\pi^0\nu_{\tau}$ decays,
respectively. As discussed above we have taken $c_4 = -0.07\pm 0.01$ and
$g_4 = -0.72 \pm 0.20$ (notice that the second process does not depend on
$g_4$). We conclude that the vector contribution ($\Gamma_V$) dominates
over the axial-vector one ($\Gamma_A$) in both channels~:
\begin{equation}
 \frac{\Gamma_A}{\Gamma_V} \, \Big|_{K \overline{K} \pi}  = \, 0.16 \pm 0.05\; , \; \; \; \;
 \frac{\Gamma_A}{\Gamma_V} \, \Big|_{K^- K^0 \pi}  = \, 0.18 \pm 0.04\;  ,  \; \; \;
\frac{\Gamma(\tau^- \rightarrow K^+ K^- \pi^- \nu_{\tau})}{\Gamma(\tau^- \rightarrow K^- K^0 \pi^0 \nu_{\tau})} \, = \, 1.4  \pm 0.3  \; ,
\end{equation}
where the errors estimate the slight variation due to the range in $c_4$ and
$g_4$. These ratios translate into a ratio of the vector current to all
contributions of $f_v = 0.86 \pm 0.04$ for the $K \overline{K} \pi$ channel
and $f_v = 0.85 \pm 0.03$ for the $K^- K^0 \pi$ one, to be compared with the
result in Ref.~\cite{Davier:2008sk}, namely $f_v(K\overline{K}\pi) = 0.20
\pm 0.03$. Our results for the relative contributions of vector and
axial-vector currents deviate strongly from most of the previous estimates,
as one can see in Table~\ref{tab:3}. Only Ref.~\cite{Gomez-Cadenas:1990uj}
pointed already to vector current dominance in these channels, although
enforcing just the leading chiral constraints and using experimental data at
higher energies.
We conclude that for all $\tau \rightarrow K K \pi \nu_{\tau}$ channels the
vector component dominates by far over the axial-vector one, though, as can
be seen in the spectra in Figs.~\ref{fig:53},\ref{fig:54}, the axial-vector
current is the dominant one in the very-low $Q^2$ regime.
\begin{table}[!ht]
\begin{center}
\begin{tabular}{|c|c|}
\hline
& \\[-2.5mm]
Source  & $\Gamma_V / \Gamma_A$  \\ [1.5mm]
\hline
& \\ [-2.5mm]
Our result &  $6  \pm 2$ \\ [1.9mm]
KS model \cite{KS5} & $0.6 - 0.7$ \\ [1.9mm]
KS model \cite{Finkemeier:1996hh} & $0.4 - 0.6$ \\ [1.9mm]
Breit-Wigner approach \cite{Gomez-Cadenas:1990uj} & $\sim 9$ \\ [1.9mm]
CVC \cite{Davier:2008sk} & $0.20 \pm 0.03$ \\ [1.9mm]
Data analysis \cite{Liu:2002mn} & $1.26 \pm 0.35$ \\[1.9mm]
\hline
\end{tabular}
\caption{\small{Comparison of the ratio of vector and axial-vector
contribution for $\tau \rightarrow KK\pi\nu_{\tau}$ partial widths.
The last two lines correspond to the $\tau^- \rightarrow K^+ K^- \pi^-
\nu_{\tau}$ process only. Results in Ref.~\cite{Finkemeier:1996hh} are an
update of Ref.~\cite{KS5}. The result of Ref.~\cite{Davier:2008sk} is
obtained by connecting the tau decay width with the CVC related $e^+ e^-
\rightarrow K_S K^{\pm} \pi^{\mp}$ (see Appendix~\ref{ap:4}). The
analysis in \cite{Liu:2002mn} was performed with a parameterization that
spoiled the chiral normalization of the form factors. }} \label{tab:3}
\end{center}
\end{table}
\par
\begin{figure}[!ht]
\begin{center}
\vspace*{0.2cm}
\includegraphics[scale=0.45,angle=-90]{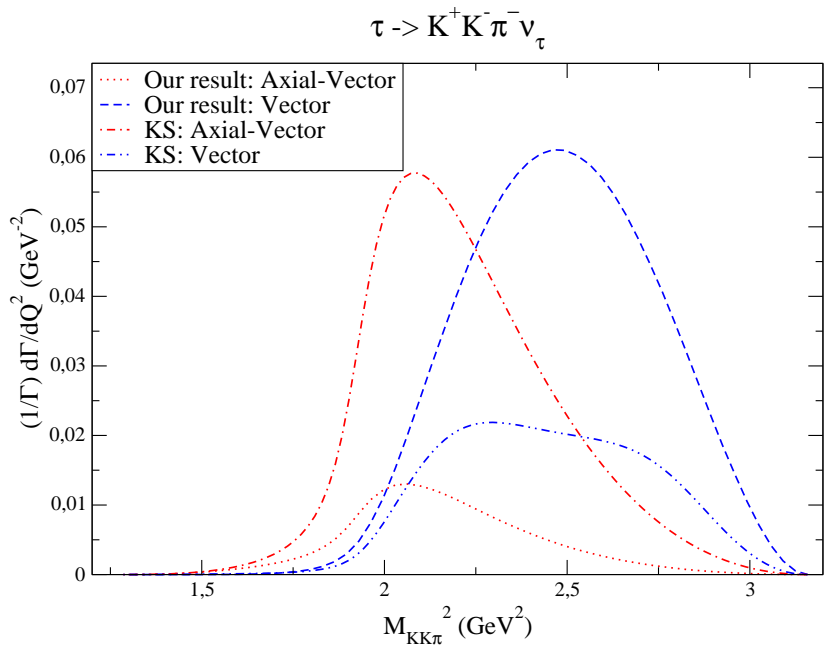}
\caption[]{\label{fig:55} \small{Comparison between the normalized $M_{K K \pi}^2$-spectra for the vector and
axial-vector contributions to the $\tau^- \rightarrow K^+ K^- \pi^- \nu_{\tau}$ channel in the KS model
\cite{Finkemeier:1996hh} and in our approach.}}
\end{center}
\end{figure}

Next we contrast our spectrum for $\tau^- \rightarrow K^+ K^- \pi^-
\nu_{\tau}$ with that one arising from the KS model worked out in
Refs.~\cite{KS5,Finkemeier:1996hh}. This comparison is by no means straight
because in these references a second and even a third multiplet of
resonances are included in the analysis. As we consider that the spectrum is
dominated by the first multiplet, in principle we could start by switching
off heavier resonances. However we notice that, in the KS model, the
$\rho(1450)$ resonance plays a crucial role in the vector contribution to
the spectrum. This feature depends strongly on the value of the $\rho(1450)$
width, which has been changed from Ref.~\cite{KS5} to
Ref.~\cite{Finkemeier:1996hh}~\footnote{Moreover within Ref.~\cite{KS5} the
authors use two different set of values for the $\rho(1450)$ mass and width,
one of them in the axial-vector current and the other in the vector one.
This appears to be somewhat misleading.}. In Fig.~\ref{fig:55} we
compare our results for the vector and axial-vector contributions with those
of the KS model as specified in Ref.~\cite{Finkemeier:1996hh} (here we have
switched off the seemingly unimportant $K^*(1410)$). As it can be seen there
are large differences in the structure of both approaches. Noticeably
there is a large shift in the peak of the vector spectrum owing to the
inclusion of the $\rho(1450)$ and $\rho(1700)$ states in the KS model
together with its strong interference with the $\rho(770)$ resonance. In our
scheme, including the lightest resonances only, the $\rho(1450)$ and
$\rho(1700)$ information has to be encoded in the values of $c_4$ and $g_4$
couplings (that we have extracted in Subsection~\ref{sect:41}) and such an
interference is not feasible. It will be a task for the experimental data to
settle this issue.
\par
\begin{figure}[!ht]
\begin{center}
\vspace*{0.2cm}
\includegraphics[scale=0.45,angle=-90]{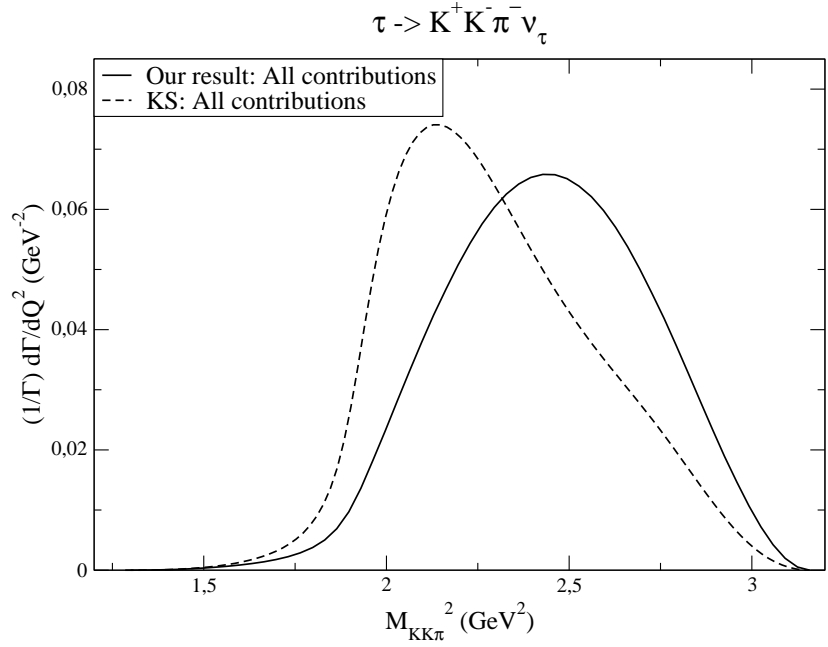}
\caption[]{\label{fig:56} \small{Comparison between the normalized $M_{K K \pi}^2$-spectra for
$\tau^- \rightarrow K^+ K^- \pi^- \nu_{\tau}$ in the KS model \cite{Finkemeier:1996hh} and in our approach.
}}
\end{center}
\end{figure}

In Fig.~\ref{fig:56} we compare the normalized full $M_{K K \pi}^2$ spectrum
for the $\tau \rightarrow K \overline{K} \pi \nu_{\tau}$ channels in the KS
model \cite{Finkemeier:1996hh} and in our scheme. The most important feature
is the large effect of the vector contribution in our case compared with the
leading role of the axial-vector part in the KS model, as can be seen in
Fig.~\ref{fig:55}. This is the main reason for the differences between
the shapes of $M_{K K \pi}^2$ spectra observed in Fig.~\ref{fig:56}.
\begin{figure}[!ht]
\begin{center}
\vspace*{0.2cm}
\includegraphics[scale=0.45,angle=-90]{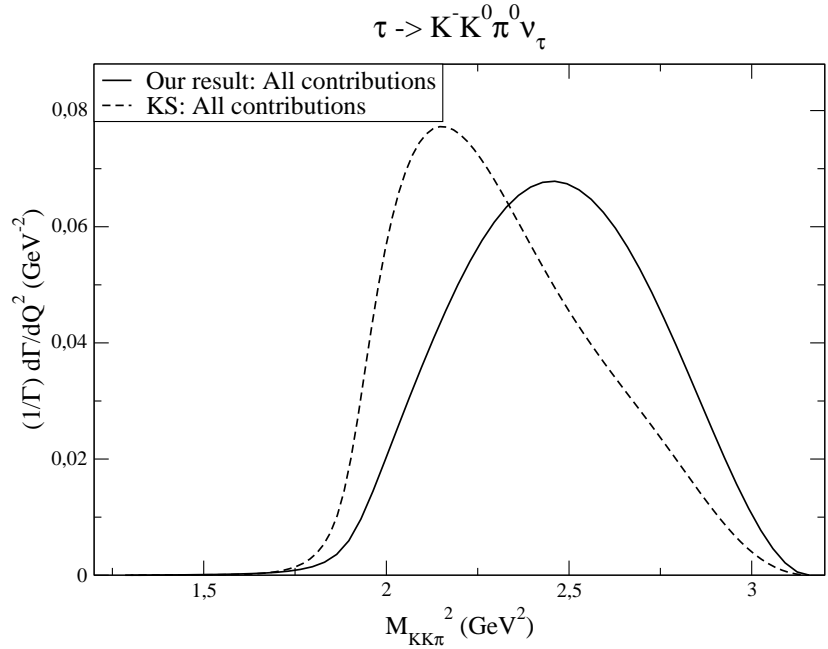}
\caption[]{\label{fig:57} \small{Comparison between the normalized $M_{K K \pi}^2$-spectra for
$\tau^- \rightarrow K^- K^0 \pi^0 \nu_{\tau}$ in the KS model \cite{Finkemeier:1996hh} and in our approach.
}}
\end{center}
\end{figure}

\section{Conclusions}

Hadron decays of the tau lepton are an all-important tool in the study of
the hadronization process of QCD currents, in a setting where resonances
play the leading role. In particular the final states of three mesons are
the simplest ones where one can test the interplay between different
resonance states. At present there are three parameterizations implemented
in the TAUOLA library to describe the hadronization process in tau decays.
Two are based on experimental data. The other alternative, namely the KS
model, though successfull in the account of the $\pi \pi \pi$ final state,
has proven to be unsuitable \cite{Liu:2002mn} when applied to the decays
into $K K \pi$ hadron states. Our procedure, guided by large $N_C$, chiral
symmetry and the asymptotic behaviour of the form factors driven by QCD, was
already employed in the analysis of $\tau \rightarrow \pi \pi \pi
\nu_{\tau}$ in Refs.~\cite{GomezDumm:2003ku} and \cite{shortp}, which
only concern the axial-vector current. Here we have applied our methodology
to the analysis of the $\tau \rightarrow K K \pi \nu_{\tau}$ channels where
the vector current may also play a significant role.
\par
We have constructed the relevant Lagrangian involving the lightest
multiplets of vector and axial-vector resonances. Then we have proceeded to
the evaluation of the vector and axial-vector currents in the large-$N_C$
limit of QCD, i.e.\ at tree level within our model. Though the widths of
resonances are a next-to-leading effect in the $1/N_C$ counting, they have
to be included into the scheme since the resonances do indeed resonate due
to the high mass of the decaying tau lepton. We have been able to
estimate the values of the relevant new parameters appearing in the
Lagrangian with the exception of two, namely the couplings $c_4$ and $g_4$,
which happen to be important in the description of $\tau \rightarrow K K \pi
\nu_{\tau}$ decays. The range of values for these couplings has been
determined from the measured widths $\Gamma(\tau^- \rightarrow K^+ K^- \pi^-
\nu_{\tau})$ and $\Gamma(\tau^- \rightarrow K^- K^0 \pi^0 \nu_{\tau})$.
\par
In this way we provide a prediction for the ---still unmeasured---
spectra of both processes. We conclude that the vector current contribution
dominates over the axial-vector current, in fair disagreement with the
corresponding conclusions from the KS model \cite{Finkemeier:1996hh} with
which we have also compared our full spectra. On the other hand, our result
is also at variance with the analysis in Ref.~\cite{Davier:2008sk}. There
are two all-important differences that come out from the comparison. First,
while in the KS model the axial-vector contribution dominates the partial
width and spectra, in our results the vector current is the one that rules
both spectrum and width. Second, the KS model points out a strong
interference between the $\rho(770)$, the $\rho(1450)$ and the $\rho(1700)$
resonances that modifies strongly the peak and shape of the $M_{KK \pi}$
distribution depending crucially on the included spectra. Not having a
second multiplet of vector resonances in our approach, we cannot provide
this feature. It seems strange to us the overwhelming role of the
$\rho(1450)$ and $\rho(1700)$ states but it is up to the experimental
measurements to settle this issue.
\par
Even if our model provides a good deal of tools for the
phenomenological analyses of observables in tau lepton decays, it may
seem that our approach is not able to carry the large amount of input
present in the KS model, as the later includes easily many multiplets of
resonances. In fact, this is not the case, since the Lagrangian can be
systematically extended to include whatever spectra of particles are needed.
If such an extension is carried out the determination of couplings
could be cumbersome or just not feasible, but, on the same footing as
the KS model, our approach would provide a parameterization to be fitted by
the experimental data. The present stage, however, has its advantages. By
including only one multiplet of resonances we have a setting where the
procedure of hadronization is controlled from the theory. This is very
satisfactory if our intention is to use these processes to learn about
QCD and not only to fit the data to parameters whose relation with the
underlying theory is unclear when not directly missing.
\par
We intend to follow our approach to analyse further relevant three
pseudoscalar channels along the lines explained in this article.

\paragraph{Acknowledgements} $\,$ \\ \label{Acknowledgements}
$\,$ \\
\hspace*{0.5cm} We wish to thank S.~Eidelman, H.~Hayashii, B.~Malaescu, O.~Shekhovtsova and Z.~Was for their
interest in this work and many useful discussions on the topic of this article.
P.~Roig has been partially supported by a FPU contract (MEC), the DFG cluster of excellence
'Origin and Structure of the Universe' and a Marie Curie ESR Contract (FLAVIAnet).
This work has been supported in part by the EU MRTN-CT-2006-035482 (FLAVIAnet),
by MEC (Spain) under grant FPA2007-60323, by the Spanish Consolider-Ingenio 2010
Programme CPAN (CSD2007-00042) and by Generalitat Valenciana under grant PROMETEO/2008/069.
This work has also been supported by CONICET and ANPCyT (Argentina), under grants PIP6009,
PIP02495, PICT04-03-25374 and PICT07-03-00818.

\appendix
\renewcommand{\theequation}{\Alph{section}.\arabic{equation}}
\renewcommand{\thetable}{\Alph{section}.\arabic{table}}
\setcounter{section}{0}
\setcounter{equation}{0}
\setcounter{table}{0}

\section{Definitions in the expressions of form factors}
\label{ap:1}

The results for the $F_1$, $F_2$ and $F_3$ form factors in $\tau \rightarrow
K K \pi \nu_{\tau}$ decays given in Eqs.~(\ref{eq:f11}), (\ref{eq:f21}),
(\ref{eq:f31}), (\ref{eq:f12}), (\ref{eq:f22}) and (\ref{eq:f32}) are
expressed in terms of the following functions~:
\begin{eqnarray} \label{eq:cde}
A^{\mbox{\tiny R}}(Q^2,x,y,m_1^2,m_2^2,m_3^2) & = &  3\, x \, + m_1^2 -m_3^2 +
\left( 1-\frac{2 G_V}{F_V} \right) \left[ 2\, Q^2-2\, x-y+m_3^2-m_2^2 \right] \, , \nonumber \\[3.5mm]
B^{\mbox{\tiny R}}(x,y,m_1^2,m_2^2) & = & 2 \, \left( m_2^2-m_1^2 \right) \,
+ \, \left( 1-\frac{2 G_V}{F_V} \right) \left[ y - x + m_1^2-m_2^2\right] \, , \nonumber \\[3.5mm]
A^{\mbox{\tiny RR}}(Q^2,x,y,m_1^2,m_2^2,m_3^2) & = & \left( \lambda' + \lambda'' \right) \,
(-3\, x + m_3^2-m_1^2)\, \nonumber \\ &&  + \, \left( 2\, Q^2+x-y+m_1^2-m_2^2 \right)
F\left( \frac{x}{Q^2}\,,\,\frac{m_2^2}{Q^2} \right) \, , \nonumber \\ [3.5mm]
B^{\mbox{\tiny RR}}(Q^2,x,y,z,m_1^2,m_2^2,m_3^2) & = & 2 \left(
\lambda'+\lambda'' \right) \left( m_1^2-m_2^2 \right)
 + \left( y-x+m_2^2-m_1^2 \right)
F\left( \frac{z}{Q^2}\,,\,\frac{m_3^2}{Q^2} \right) \, , \nonumber \\ [3.5mm]
 C^{\mbox{\tiny R}}(Q^2,x,m_1^2,m_2^2,m_3^2) & = &
 (c_1-c_2+c_5) \, Q^2 - ( c_1-c_2-c_5+2 c_6) \, x \nonumber \\
 & & \, + (c_1+c_2 + 8 c_3 -c_5) \, m_3^2 + 8\, c_4\, (m_1^2-m_2^2) \, , \nonumber \\[3mm]
C^{\mbox{\tiny RR}}(Q^2,x,m^2) & = & d_3 \, (Q^2+x)+ (d_1+8\, d_2-d_3) \, m^2 \, , \nonumber \\ [3.5mm]
D^{\mbox{\tiny R}}(Q^2,x,y) & = & (g_1+2 \, g_2-g_3)\, (x+y) -2 \,g_2 \, (Q^2+m_K^2)
\nonumber \\
&&  - (g_1-g_3)\, ( 3\,m_K^2+m_{\pi}^2 ) +2 \, g_4 \, (m_K^2+m_{\pi}^2) +2 \, g_5 \, m_K^2 \, , \nonumber \\[3mm]
E^{\mbox{\tiny R}}(x,y) & = & (g_1+2 \, g_2-g_3)\, (x-y) \, .
\end{eqnarray}
Here $u = Q^2 - s - t + m_1^2 + m_2^2 + m_3^2$ and $F(x,y) = \lambda'' +
\lambda'\, x - \lambda_0\, y$, where $\lambda_0$, $\lambda'$ and $\lambda''$
are combinations of the $\lambda_i$ couplings defined in
Eq.~(\ref{eq:lag21})~:
\begin{eqnarray}
\label{eq:lambdias}
 - \sqrt{2} \, \lambda_0 & = & 4 \, \lambda_1 + \lambda_2 + \frac{\lambda_4}{2} + \lambda_5 \, ,\nonumber \\
\sqrt{2} \, \lambda' & = & \lambda_2 - \lambda_3+ \frac{\lambda_4}{2} + \lambda_5 \, , \\
\sqrt{2} \, \lambda'' & = & \lambda_2 - \frac{\lambda_4}{2} - \lambda_5 \, . \nonumber
\end{eqnarray}

\appendix
\renewcommand{\theequation}{\Alph{section}.\arabic{equation}}
\renewcommand{\thetable}{\Alph{section}.\arabic{table}}
\setcounter{section}{1}
\setcounter{equation}{0}
\setcounter{table}{0}

\section{$2 \, g_4 + g_5$ from $\omega \rightarrow \pi^+ \pi^- \pi^0$}
\label{ap:3}

The process
$
\omega \rightarrow \pi^{+}(k_1)\, \pi^{-}(k_2) \, \pi^{0}(k_3)
$
provides us with an estimate for the combination of couplings $2 \, g_4+ g_5$.
We will denote the polarization vector of the $\omega$ as $\varepsilon_\omega^\sigma$ and use the kinematic invariants $s_{ij}\,=\,(k_i\,+\,k_j)^2$.
\par
The amplitude for this process has two contributions. The first one,
mediated by the $\rho(770)$ resonance was already studied in
Ref.~\cite{RuizFemenia:2003hm}, where it was concluded that the contribution
of a pure local amplitude was necessary to fulfill the phenomenological
determinations. This piece can be obtained from our Lagrangian in
Eq.~(\ref{eq:omegappp}). The full result is given by~:
\begin{eqnarray}
i\,\mathcal{M}_{\omega \rightarrow \pi^+ \pi^- \pi^0} & =  & i\,
\varepsilon_{\alpha\beta\rho\sigma}k_1^\alpha k_2^\beta k_3^\rho \varepsilon_\omega^\sigma\, \,  \frac{8 \, G_V}{M_\omega F^3} \, \times \nonumber \\ [3mm]
&& \times \Bigg\lbrace \frac{m_\pi^2(d_1\,+\,8\,d_2\,-\,d_3)+(M_\omega^2\,+\,s_{12})d_3}{M_\rho^2\,-\,s_{12}}
  + \left\lbrace s_{12}\to s_{13}\right\rbrace + \left\lbrace s_{12}\to s_{23}\right\rbrace \nonumber \\ [3mm]
&& \; \; \; \; \;  \left. + \, \frac{ \sqrt{2}}{G_V \,  M_V}\left[ (g_1\,-\,g_2\,-\,g_3)(M_\omega^2\,-\,3\,m_\pi^2)\,+\,3\,m_\pi^2(2\,g_4\,+\,g_5)\right] \right\rbrace\,,
\end{eqnarray}
where we have assumed ideal mixing between the states $| \, \omega_8 \rangle$ and $| \, \omega_1 \rangle$~:
\begin{equation}
 |\,  \omega \rangle\,=\,\sqrt{\frac{2}{3}}  |  \, \omega_1\rangle+\sqrt{\frac{1}{3}}  | \,  \omega_8\rangle \,.
\end{equation}
Using the experimental figure for $\mbox{BR}(\omega \rightarrow \pi^+ \pi^-
\pi^0) = 0.892 \pm 0.007$ \cite{PDG2008}, introducing the already known
combinations of couplings as discussed in Sect.~\ref{sect:4} and taking $G_V
= F_V/F^2$ with $F_V$ given by Eq.~(\ref{eq:set1}) we find~:
\begin{equation}
\label{eq:gii}
 2 \, g_4 + g_5 \, = \, -0.60 \pm 0.02 \, .
\end{equation}
We will use this result to eliminate $g_5$ in terms of $g_4$, that remains unknown.

\appendix
\renewcommand{\theequation}{\Alph{section}.\arabic{equation}}
\renewcommand{\thetable}{\Alph{section}.\arabic{table}}
\setcounter{section}{2}
\setcounter{equation}{0}
\setcounter{table}{0}

\section{Relation between $\sigma \left( e^+ e^- \rightarrow KK \pi \right)$ and $\Gamma \left( \tau \rightarrow K K \pi \nu_{\tau} \right)$}
\label{ap:4}

Using $SU(2)_I$ symmetry, one can derive several relations between exclusive isovector hadron modes produced in $e^+e^-$ collisions and those related with the vector current ($F_3$ form factor) in $\tau$ decays. In
particular we find~:
\begin{equation}\label{eq:CVC1}
 \frac{\mathrm{d}}{\mathrm{d}Q^2} \,
\Gamma\left( \tau^- \to K^0 K^- \pi^0 \nu_\tau\right) \Bigg|_{F_3} \, = \, f(Q^2)\;\sigma_{I=1} \left(e^+e^-\to K^-K^0\pi^+\right)\,,
\end{equation}
where
\begin{equation} \label{eq:fQ2}
f(Q^2)=\frac{G_F^2|V_{ud}|^2}{128(2\pi)^5 M_\tau}\,\left(\frac{M_\tau^2}{Q^2}-1 \right)^2\,\frac{1}{3}\,\left( 1+\frac{2Q^2}{M_\tau^2}\right)\,\left( \frac{96\pi}{\alpha^2} \right)\,Q^6\,.
\end{equation}
Analogously, one can also derive:
\begin{eqnarray}\label{eq:rels_iso_2}
 2 \, \frac{\mathrm{d}}{\mathrm{d}Q^2} \, \Gamma\left( \tau\to K^+ K^- \pi^- \nu_\tau\right) \Bigg|_{F_3} & = & f(Q^2)\,\left[ \sigma_{I=1} \left(e^+e^-\to K^+\overline{K}^0\pi^-\right)\, \right. \\ \nonumber
 && \; \; \; \; \; \; \; \; \; \; \; \;
   +\,2 \, \sigma_{I=1} \left(e^+e^-\to K^+\,K^-\pi^0\right)\Big] \,.
\end{eqnarray}
Summing these equations one obtains~:
\begin{equation}\label{eq:rels_iso_all}
\sum_{i=1}^3 \frac{\mathrm{d}}{\mathrm{d}Q^2} \, \Gamma\left( \tau\to (K K \pi)_i \nu_\tau\right) \Bigg|_{F_3} \,  = \, f(Q^2)\,\sum_{i=1}^4\,\sigma_{I=1} \left(e^+e^-\to (K K \pi)_i \right)\,,
\end{equation}
where the sums run over all possible charge channels in each case. If
isovector and isoscalar components were splitted for all channels
Eq.~(\ref{eq:rels_iso_all}) would allow us to fit the data using our vector
form factors for $\tau\to (K K \pi)_i \nu_\tau$.
\par
BABAR has managed to split the isoscalar and isovector components in the
cross sections $\sigma\left(e^+e^-\to K_S K^{\pm} \pi^{\mp}\right)$
\cite{Aubert:2007ym}. The $I=1$ component of $\sigma\left(e^+e^-\to K K\,
\pi\right)$ needs to be used, under the hypothesis of CVC, to obtain the
spectral function of the processes $\tau \to KK\pi\nu_\tau$, and thus to
help the extraction of $\alpha_S(M_\tau)$ \cite{Davier:2008sk}. However, it
is not straightforward to obtain the inclusive $I=1$ component of
$\sigma\left(e^+e^-\to K K \pi\right)$ from the measured value of
$\sigma\left(e^+e^-\to K_S^0 K^{\pm} \pi^{\mp}\right)$. In fact, using only
$SU(2)_I$ symmetry this is not possible. There are two further assumptions
that need to be done in order to obtain the expression usually employed~:
\begin{equation} \label{eq:factor3}
\sigma\left(e^+e^-\to K K \pi\right) \, = \, 3 \; \sigma\left(e^+e^-\to K_S K^{\pm} \pi^{\mp}\right) \, .
\end{equation}
The first one is to assume that the processes $e^+e^-\to KK\pi$ are
dominated by $K^*$-exchange. According to recent Dalitz plot analyses,
\cite{Aubert:2007ym} this is indeed a good approximation. However,
$SU(2)_I$ symmetry and $K^*$ dominance do not allow to relate
$\sigma\left(e^+e^-\to K_S K^{-} \pi^{+}\right)$ and $\sigma\left(e^+e^-\to
K^+ K^{-} \pi^{0}\right)$ as given by Eq.~(\ref{eq:factor3}). Under $K^*$
dominance there are two intermediate chains that account for each final
state~:
\begin{eqnarray} \label{eq:interfiero}
  A \left( K^+ K^- \pi^0 \right)  & = &
 A \left( e^+e^-\to(K^{*-})\,K^+\to K^- \pi^{0}\, K^+ \right) \nonumber \\
& & +  A \left( e^+e^-\to(K^{*+})\, K^-\to K^+ \pi^{0}\, K^- \right) \;  \equiv \, B \, + \, C \, ,
\nonumber \\ \nonumber \\
  A \left( K^- K^0 \pi^+ \right)  & = &
  A \left( e^+e^-\to(K^{*+})\,K^-\to K^0 \pi^{+} \,K^{-} \right)  \nonumber \\
& & +   A \left( e^+e^-\to(\overline K^{*0})\, K^0\to \pi^+ K^- \, K^0 \right) \; \equiv \,
\sqrt{2} \, \left( B \, - \, C \right) \, .
\end{eqnarray}
Accordingly we conclude that the relation $\sigma\left(e^+e^-\to K^0 K^{-}
\pi^{+}\right) = 2 \; \sigma\left(e^+e^-\to K^+ K^{-} \pi^{0}\right)$,
which is necessary to derive Eq.~(\ref{eq:factor3}), can only be obtained
by neglecting the interference terms arising from
Eqs.~(\ref{eq:interfiero}).
\par
We have checked the accuracy of this assumption using two different
parameterizations for the involved hadronization amplitudes. First, we have
employed a parameterization following the KS-like model used for tau decays
into $KK\pi$ modes, using the values in Ref.~\cite{Finkemeier:1996hh}.
Furthermore, we have used our expressions obtained within R$\chi$T in
Sect.~\ref{sect:3}. In both cases, we have not set the contributions of
other resonances than the $K^*$ to zero, although we have checked that they
are of very little importance. With both kinds of
parameterizations either at $\Gamma\left(\tau\to KK\pi \nu_\tau\right)$ or
at $\sigma\left(e^+e^-\to KK\pi\right)$ the error of assuming that
interference effects are negligible is at least of order 30 $\%$ in
$e^+e^-$, being even larger in $\tau$ decays.
\begin{table}[ht]
\begin{center}
\renewcommand{\arraystretch}{1.2}
\begin{tabular}{|c|c|}
\hline
Parameterization & $\sigma\left(e^+e^-\to K^0 K^{-} \pi^{+}\right)/ 2\, \sigma\left(e^+e^-\to K^+ K^{-} \pi^{0}\right)$ \\
\hline
\cite{Finkemeier:1996hh}& $0.74$  \\
R$\chi$T& $0.36$ \\
\hline
\end{tabular}
\caption{\small{Check of the validity of relation (\ref{eq:factor3}) -and thus of neglecting interference effects- for different hadronization parameterizations.
The $Q^2$-endpoint in the cross-sections has been taken at~$\sim2$ GeV$^2$. We estimate the error of the
R$\chi$T prediction to be around $30 \, \%$.}}
\label{tab:checke+e-}
\end{center}
\end{table}
In the same way, taking for granted the rightness of the very accurate
assumptions of $SU(2)_I$ and $K^*$ dominance in $\tau\to KK\pi \nu_\tau$, it
is still not possible to relate the widths to $K^+ K^- \pi^-$ and $K^- K^0
\pi^0$ in a model independent way. Under the hypothesis of negligible
interference one would obtain
\begin{equation} \label{eq:factor1}
2 \; \Gamma\left(\tau\to K^+K^-\pi^- \nu_\tau\right)=\Gamma\left(\tau\to K^-K^0\pi^0 \nu_\tau\right) \,.
\end{equation}
As can be observed in Table~\ref{tab:checktau}, within the above
considered models this relation does not hold.
\begin{table}[ht]
\begin{center}
\renewcommand{\arraystretch}{1.2}
\begin{tabular}{|c|c|}
\hline
Parameterization & $\Gamma\left(\tau\to K^- K^{0} \pi^{0}\right)/ 2 \, \Gamma\left(\tau\to K^+ K^{-} \pi^{-}\right)$\\
\hline
\cite{Finkemeier:1996hh}& $0.62$\\
R$\chi$T& $0.36$\\
\hline
\end{tabular}
\caption{\small{Check of the validity of relation (\ref{eq:factor1}) -and thus of neglecting interference effects- for different hadronization parameterizations.  We estimate the error of the
R$\chi$T prediction to be around $30 \, \%$.}}
\label{tab:checktau}
\end{center}
\end{table}

\appendix
\renewcommand{\theequation}{\Alph{section}.\arabic{equation}}
\renewcommand{\thetable}{\Alph{section}.\arabic{table}}
\setcounter{section}{3}
\setcounter{equation}{0}
\setcounter{table}{0}

\section{Numerical input}
\label{ap:input}

For the numerics we use, if nothing is specified, the masses given in Ref.~\cite{PDG2008}.
From the analyses of Refs.~\cite{shortp,Jamin:2008qg} we use, as input, the following
numerical values of the parameters that appear in our study~:
\begin{eqnarray} \label{eq:set1}
F \, = \, 0.0924 \, \mbox{GeV} \; & \; , \;  & \;  F_V \, = \, 0.180 \, \mbox{GeV} \; \; \; \; \; \; \;  , \; \; \; \; 
F_A \, = \, 0.149 \, \mbox{GeV} \;,  \nonumber \\
M_V \, = \, 0.775 \, \mbox{GeV}  \; & \; , \;  & \; 
M_{K^*} \, = \, 0.8953 \, \mbox{GeV}\;  \; \; , \; \; \; \; 
M_{a_1} \, = \, 1.120 \, \mbox{GeV}  \, .
\end{eqnarray}
Then we get $\lambda'$, $\lambda''$ and $\lambda_0$ from the first equalities in Eq.~(\ref{eq:lambres}). Incidentally
we can also determine the value of $M_A \simeq 0.91 \, \mbox{GeV}$. Notice that this value for $M_A$ is
slightly lower than the result obtained in Ref.~\cite{Mateu:2007tr}. Our preferred set of values in 
Eq.~(\ref{eq:set1}) satisfies reasonably well all the short-distance constraints pointed out in 
Sect.~\ref{sect:4}, with a deviation from Weinberg sum rules of at most $10 \%$, perfectly compatible
with deviations due to the single resonance approximation.

\end{document}